\documentclass[twocolumn]{aastex701}
\usepackage{amsmath}

\newcommand{\sjtuphytdli}{State Key Laboratory of Dark Matter Physics, Tsung-Dao Lee Institute \& School of Physics and Astronomy, Shanghai Jiao Tong University, Shanghai 200240, China}

\newcommand{\sjtuphy}{Shanghai Key Laboratory for Particle Physics and Cosmology, and Key Laboratory for Particle Physics, Astrophysics and Cosmology, Ministry of Education, Shanghai Jiao Tong University, Shanghai 200240, China}

\usepackage{xcolor}
\definecolor{red}{rgb}{1,0,0}

\begin{document}

\title{The Universality of Halo Growth and Merger Rates in Modified Gravity Theories}

\correspondingauthor{Zhongxu Zhai, Jiaxin Han}
\email{zhongxuzhai@sjtu.edu.cn, jiaxin.han@sjtu.edu.cn}

\author{Haicong Xu}
\affiliation{\sjtuphytdli}
\affiliation{\sjtuphy}
\email{haicongxu@sjtu.edu.cn}

\author{Yechun Yong}
\affiliation{\sjtuphytdli}
\email{yongyechun@sjtu.edu.cn }

\author{Feihong He}
\affiliation{Kavli Institute for Astronomy and Astrophysics, Peking University, Beijing 100871, China;}
\email{feihong.he@pku.edu.cn}

\author{Zhongxu Zhai}
\affiliation{\sjtuphytdli}
\affiliation{\sjtuphy}
\affiliation{Waterloo Center for Astrophysics, University of Waterloo, Waterloo, ON N2L 3G1, Canada}
\affiliation{Department of Physics and Astronomy, University of Waterloo, Waterloo, ON N2L 3G1, Canada}
\email{zhongxuzhai@sjtu.edu.cn}

\author{Jiaxin Han}
\affiliation{\sjtuphytdli}
\affiliation{\sjtuphy}
\email{jiaxin.han@sjtu.edu.cn}

\author{Baojiu Li}
\affiliation{Institute for Computational Cosmology, Department of Physics, Durham University, South Road, Durham DH1 3LE, UK}
\email{baojiu.li@durham.ac.uk}

\begin{abstract}

In the standard $\Lambda$CDM framework, the merger rate of dark matter haloes has been found to follow a nearly universal form, with only weak dependence on halo mass and redshift. We examine whether this universality is preserved in modified gravity (MG) models, where the growth of cosmic structure can deviate from that of general relativity (GR). We use the \textsc{ELEPHANT} suite of large $N$-body simulations to study halo merger rates and mass accretion histories in different modified gravity models, including two representative scenarios beyond-GR: the Hu--Sawicki chameleon $f(R)$ model and the normal branch of the Dvali--Gabadadze--Porrati model. 
For the mass accretion histories, the modified gravitational effect can produce systematic departures from GR, generally corresponding to more rapid halo growth. The deviations are relatively small when the MG effect is weak, while they become more pronounced in models with strong deviation from GR and can exceed $15\%$ in the extreme case.

For the halo merger rate, we find that the nearly universal behavior is well preserved across all the gravity models, with only weak dependence on redshift and halo mass, reflecting the unbiased accretion nature of halo growth. However, apparent deviation from this universality can arise if the halo mass is not defined consistently with the modified gravity model. 

\end{abstract}


\section{Introduction} \label{sec:intro}

In the standard $\Lambda$CDM cosmology, dark matter halos form from small perturbations in the primordial density field of the Universe and grow hierarchically through the accretion of diffuse matter and mergers with one another.
These halos provide the gravitational potential wells in which baryons can accumulate and eventually form galaxies. The resulting distributions of halos and galaxies are closely coupled to cosmological evolution and can therefore be used to constrain the expansion history of the Universe \citep{Eisenstein2005, Cole2005, Dawson2013, Dawson2016, AbdulKarim2025, Adame2025}. On the other hand, observations of distant Type Ia supernovae and other cosmological probes have shown that the Universe entered a phase of accelerated expansion at late times \citep{Riess1998, Perlmutter1999}.

The origin of this late-time acceleration remains an important question in modern cosmology.
Within the framework of general relativity (GR), it is commonly attributed to a cosmological constant or dark energy \citep{Copeland2006}, which may be interpreted as an exotic component with negative pressure.
Alternatively, late-time cosmic acceleration may signal a breakdown of GR on cosmological scales and can be explained by modifying gravity (MG) without introducing an additional dark energy component \citep{Clifton2012}.

To satisfy stringent observational constraints on gravity, particularly those under which GR has already been precisely tested, modified gravity models typically invoke screening mechanisms \citep{Vainshtein1972, Khoury2004}. These mechanisms ensure that deviations from GR are suppressed in high-density environments, such as the Solar System. On these scales, MG models converge towards GR.
However, on cosmological scales, the formation and evolution of large-scale structure can exhibit significant deviations from GR.
Studying these signatures therefore provides a powerful probe of gravitational effects and an important test of GR \citep{Koyama2016, Hernandez-Aguayo2018, Cautun2018, Armijo2018, Ishak2019, Paillas2019, Hernandez-Aguayo2019, Armijo2026}.

To this end, a variety of methods have been proposed and investigated to search for signatures of MG effects in large-scale structure.
For example, studies of the halo mass function have revealed a nearly universal form that remains approximately independent of redshift in modified gravity models \citep{Gupta2022}. However, the velocity field can differ between MG models and GR. Previous work adopted an empirical velocity-bias model and showed that deviations from GR can have a noticeable impact on the galaxy correlation function on small to intermediate scales \citep{Wang2025}.
Cosmic voids, whose low-density environments are less affected by screening mechanisms, have been widely explored as sensitive probes of modified gravity, with several studies investigating optimal void-finding methods to maximize the MG signal \citep{Cautun2018, Paillas2019}.
In addition, marked statistics have been shown to efficiently distinguish MG models from GR by emphasizing environmentally dependent effects \citep{ Hernandez-Aguayo2018, Armijo2018, Armijo2026}, while redshift-space distortion measurements provide complementary constraints on the growth of structure and gravity \citep{Hernandez-Aguayo2019}.
We note that most previous studies have mainly focused on the abundance, clustering \citep{Arnalte-Mur2017}, and environmental dependence of halos and galaxies, while paying relatively less attention to halo merger rates and mass accretion histories, which can play an important role in building a more accurate description of the underlying theories of gravity.

Within the $\Lambda$CDM framework, a series of studies have investigated the merger processes of dark matter halos and their accretion histories using merger trees constructed from cosmological simulations \citep{Fakhouri2008, Fakhouri2009,  Genel2009, Stewart2009, McBride2009, Wetzel2009, Fakhouri2010, Genel2010, Poole2017, Dong2022, Jiang2025, Jiang2026}.
Many of these studies have also proposed empirical fitting formulae for halo merger rates and accretion histories.
Despite differences in their detailed parametrizations, these formulae commonly capture the dependence of the merger rate on descendant halo mass, merger mass ratio, and redshift.

In particular, \citet{Fakhouri2010} found that the halo merger rate can be well described by a simple fitting formula consisting of a power law at the low-mass-ratio end and an exponential decay at the high-mass-ratio end.
The dependence on descendant halo mass and redshift is weak, approximately scaling as $M_h^{0.13}$ and $(1+z)^{0.1}$, respectively.
This weak dependence implies a high degree of universality in the merger rate.

Another study by \citet{Dong2022} found that a specific form of the merger rate, defined as $dN_{\rm merger}(\xi| M, z)/d\xi/d\log M(z)$, exhibits similar universal behaviour.
In this formulation, the merger rate depends only on the mass ratio of the merging halo pair, $\xi$, and shows no dependence on the host halo mass $M$ or redshift $z$ over a wide range of halo masses ($10^{12} \lesssim M \lesssim 10^{14}, h^{-1} {\rm M_\odot}$) and merger ratios ($\xi \gtrsim 10^{-2}$).
More importantly, this universal relation has been tested in simulations with different values of $\Omega_m$ and $\sigma_8$ and was found to remain unchanged. This result suggests that, once the overall mass growth of the host halo is accounted for, halos accrete progenitors with a nearly invariant distribution of merger mass ratios.
This universality reveals a strong self-similarity in the merger-driven growth of halos: the merger rate scales with the mass growth rate of the host halo, while the mass-ratio distribution of accreted progenitors remains nearly the same at each step of halo growth \citep{Dong2022, Jiang2025, Jiang2026}. This self-similarity can be naturally understood within the unbiased accretion framework \citep{Han16}, which posits that haloes grow through the indiscriminate accretion of diffuse matter and surrounding haloes. 

If this self-similarity reflects a generic feature of hierarchical gravitational clustering, it is natural to ask whether such universal behaviour persists in modified gravity scenarios and whether it remains universal across different modified gravity models despite their distinct deviations from GR. In this work, we focus on this question using a suite of large $N$-body simulations of modified gravity. We consider two representative models: the $f(R)$ gravity model \citep{Sotiriou2010, DeFelice2010} and the normal branch of the Dvali–Gabadadze–Porrati (nDGP) model \citep{Dvali2000}. Based on the simulation data, we construct halo merger trees and then measure the mass accretion histories (MAHs) and halo merger rates to examine whether their universal behaviour is preserved across different gravity models.

This paper is organized as follows. In Section~\ref{sec:simulation}, we introduce the modified gravity models, the simulation data, and the \textsc{HBT+} algorithm.
In Section~\ref{subsec:merger_rates_mah}, we describe the methods used to extract halo merger rates and MAHs.
We present the main results in Section~\ref{sec:results}, discuss these results in Section~\ref{sec:discussion}, and summarize our conclusions in Section~\ref{sec:conclusion}.

\section{Simulation}
\label{sec:simulation}

In this section, we briefly introduce the two modified gravity models considered in this work, namely the $f(R)$ gravity model and the DGP model. 
We then describe the $N$-body simulations and the \textsc{HBT+} subhalo finder.

\subsection{Modified Gravity Models}
\label{subsec:mg_models}

\subsubsection{\texorpdfstring{$f(R)$}{f(R)} Gravity}
\label{subsubsec:fr_gravity}

In $f(R)$ gravity, the Ricci scalar $R$ in the Einstein-Hilbert action is replaced by $R+f(R)$. The action therefore becomes
\begin{equation}
    S = \frac{1}{16\pi G}\int d^4x \sqrt{-g}\, \left[R+f(R)\right] + S_{\rm M}.
    \label{eq:fr_action}
\end{equation}
Varying this action yields the modified Einstein equation,
\begin{equation}
    G_{\mu\nu}
    + f_R R_{\mu\nu}
    - g_{\mu\nu}\left(\frac{1}{2}f - \Box f_R\right)
    - \nabla_\mu \nabla_\nu f_R
    = 8\pi G T^{\rm m}_{\mu\nu},
    \label{eq:fr_field_equation}
\end{equation}
where $G_{\mu\nu}$ is the Einstein tensor, $R_{\mu\nu}$ is the Ricci tensor, $T^{\rm m}_{\mu\nu}$ is the matter energy-momentum tensor, and $f_R \equiv df(R)/dR$ is the scalar degree of freedom introduced by the modification of gravity. Here $\nabla_\mu$ denotes the covariant derivative, and $\Box \equiv \nabla^\mu \nabla_\mu$ is the d'Alembert operator.

Several functional forms of $f(R)$ are possible, each of which determines the detailed behavior of the modified gravitational force.
In this work, we adopt the Hu–Sawicki model \citep{Hu2007}, in which $f(R)$ is given by
\begin{equation}
    f(R) = -m^2
    \frac{c_1\left(-R/m^2\right)^n}
    {c_2\left(-R/m^2\right)^n + 1},
    \label{eq:hs_fr_model}
\end{equation}
where $m^2 \equiv H_0^2 \Omega_{\rm m}$.
$H_0$ is the present-day Hubble parameter, $\Omega_{\rm m}$ is the present-day matter density parameter, and $c_1$, $c_2$, and $n$ are free parameters of the model.

In practice, the parameter combination $c_1/c_2^2$ is usually replaced by $f_{R0}$, which denotes the present-day value of $f_R$. 
The relation between them is
\begin{equation}
    \frac{c_1}{c_2^2}
    =
    -\frac{1}{n}
    \left[
        3\left(
        1 + 4\frac{\Omega_\Lambda}{\Omega_{\rm m}}
        \right)
    \right]^{n+1}
    f_{R0}.
    \label{eq:c1_c2_fr0_relation}
\end{equation}
In this work, we use three $f(R)$ simulations with $n=1$ and $f_{R0}=-10^{-6}$, $-10^{-5}$, and $-10^{-4}$, hereafter referred to as F6, F5, and F4, respectively. In this parameterization, F4 exhibits the strongest deviation from GR, F6 the weakest, and F5 an intermediate deviation.
\subsubsection{DGP Model}
\label{subsubsec:dgp_model}

The Dvali-Gabadadze-Porrati (DGP) model \citep{Dvali2000, Maartens2010} is a braneworld gravity model in which our four-dimensional Universe is embedded in a five-dimensional bulk spacetime. 
In this work, we focus on the normal branch of the DGP model (nDGP), which requires an additional dark-energy component to reproduce a $\Lambda$CDM-like expansion history. 
The background expansion can be written as 
\begin{equation}
\frac{H(a)}{H_0}=\sqrt{\Omega_{\rm m0}a^{-3}+\Omega_{\rm DE}(a)+\Omega_{r_{\rm c}}}-\sqrt{\Omega_{r_{\rm c}}},
\end{equation}
where $\Omega_{r_{\rm c}} = 1/(4H_0^2 r_{\rm c}^2)$.
The deviation from GR is characterized by the crossover-scale parameter $H_0 r_{\rm c}$, with larger values corresponding to weaker deviations. The GR limit is recovered as $H_0 r_{\rm c}\rightarrow \infty$.
In addition, nDGP recovers GR in high-density environments through the Vainshtein screening mechanism, which suppresses the fifth force on nonlinear scales and allows the model to satisfy local gravity constraints \citep{Vainshtein1972}.
In this work, we use two nDGP simulations with $H_0 r_{\rm c}=5$ and $H_0 r_{\rm c}=1$, hereafter referred to as N5 and N1, respectively.

\subsection{Simulation and halo catalog}
\label{subsec:nbody_simulations}

We use the \textsc{ELEPHANT} (Extended LEnsing PHysics using ANalytic ray Tracing) suite of dark-matter-only modified gravity $N$-body simulations \citep{Cautun2018, Paillas2019, Hernandez-Aguayo2019}. The suite includes one GR simulation, three $f(R)$ simulations corresponding to F4, F5, and F6, and two nDGP simulations corresponding to N1 and N5.
The $f(R)$ simulations were performed using the \textsc{ECOSMOG} code \citep{Li2012b}, while the nDGP simulations were run using its modified version, \textsc{ECOSMOG-V} \citep{Li2013}.
The simulations adopt the following cosmological parameters: ${\Omega_{\rm b}, \Omega_{\rm CDM}, h, n_{\rm s}, \sigma_8} = {0.046, 0.235, 0.697, 0.971, 0.82}$.

Each simulation follows the evolution of $N_{\rm p}=1024^3$ dark matter particles in a cubic box with a side length of $L_{\rm box}=1024 h^{-1}{\rm Mpc}$ and a particle mass of $m_{\rm p}=7.798\times10^{10}h^{-1}{\rm M_\odot}$.
The initial conditions were generated at $z_{\rm ini}=49$ using the \textsc{MPGRAFIC} code \citep{Prunet2008}.
The simulations for all gravity models start from the same initial density field, since the effects of modified gravity on large-scale structure formation are negligible at such an early time \citep{Cautun2018}. This also ensures that comparisons between different models are not affected by cosmic variance.

With each simulation snapshot, we first construct halo catalogues using the friends-of-friends (FoF) algorithm, with a linking length equal to 0.2 times the mean particle separation.
These halo catalogues, together with the particle data from the simulation, are then provided to the \textsc{HBT+} code \citep{Han2012, Han2018} to identify and track self-bound substructures within halos and to construct merger trees of subhalos.

We adopt the FoF halo mass, defined as the total mass of all particles in a FoF group, as our default halo mass definition. Unless otherwise stated, all results presented in Section \ref{sec:results} adopt this halo mass definition.

\subsection{Subhalo catalog and merger tree}
\label{subsec:hbt}
We adopt \textsc{HBT+} to identify subhalos and build merger trees for our analysis.
\textsc{HBT+} (Hierarchical Bound Tracing) \citep{Han2012, Han2018} is a history-based subhalo finder designed to identify and trace the evolution of subhalos across simulation snapshots. 
Unlike conventional subhalo finders that identify structures independently at each snapshot, \textsc{HBT+} follows the bound particle set of each subhalo through time \citep{Han2012}, allowing it to robustly track mergers and tidal stripping.

Starting from the FoF halo catalogue, \textsc{HBT+} identifies gravitationally bound particle as source sets and updates them across snapshots. \textsc{HBT+} tracks the hierarchical merger events of subhalos and constructs subhalo merger trees while preserving particle continuity. Each subhalo is assigned a unique track, which naturally defines its progenitor and descendant through the evolutionary history. Within a given FoF halo, the most massive subhalo is defined as the central subhalo. This history-based approach provides a robust and physically consistent description of halo assembly and merger histories. We refer the readers to \cite{Han2012} and \cite{Han2018} for more details.

\section{Extracting MAHs and Merger Rates}
\label{subsec:merger_rates_mah}

For the MAHs, we start from a given halo at $z=0$ and trace its main branch backward through the merger tree by identifying the most massive progenitor at each earlier snapshot. 
The resulting mass evolution along the main branch defines the MAHs, $M(z)$. In our halo catalogues, each FoF halo is assigned a descendant halo in the next snapshot. 
A descendant halo with mass $M_0$ at redshift $z_{\rm d}$ can be traced back to $N_{\rm p}$ progenitor haloes at the previous snapshot, $z_{\rm p}=z_{\rm d}+\Delta z$.
We rank the progenitors by mass, $M_1 \geq M_2 \geq \cdots \geq M_{N_{\rm p}}$, where $M_1$ is the mass of the most massive progenitor. 
For each secondary progenitor, we define the progenitor mass ratio as $\xi_i \equiv M_i/M_1$, with $i=2,\cdots,N_{\rm p}$. Note our definition of the main progenitor differs from the default \text{HBT+} choice, which defines the main progenitor according to not only mass but also kinematic distance. Our mass-based choice has the preferential property of $\xi<1$ by construction.

To reduce the impact of numerical resolution on our results, we only include well-resolved halos in the halo catalogues. 
We require descendant halos to contain at least 200 particles, corresponding to a minimum halo mass of $10^{13.2}\,h^{-1}{\rm M_\odot}$. 
For progenitor halos, we impose a lower limit of 45 particles, and only mergers involving progenitors above this threshold are regarded as effective merger events in our analysis. As the merger rate is expected to be largely independent on the host halo mass but only depends on the mass ratio~\citep{Dong2022}, these cuts are not expected to bias our measurements. In addition, for most of the analysis we bin halos by mass, and only present our results above the corresponding mass ratio limit in each bin which is free from any potential selection bias.

Following \citet{Dong2022}, we measure the specific merger rate, defined as
\begin{equation}
    \frac{dN_{\rm merge}(\xi|M,z)}{d\xi\,d\log M}
    =
    \frac{dN_{\rm merge}/dz/d\xi}{d\log M/dz},
    \label{eq:specific_merger_rate}
\end{equation}
where $N_{\rm merge}$ is the number of merger events and $M$ denotes the halo mass. 
This quantity measures the mean number of merger events, binned by progenitor mass ratio, normalized by the logarithmic growth of the host halo mass.

In practice, for descendant haloes in a given mass bin at redshift $z_{\rm d}$, we count the effective merger events occurring between two neighbouring snapshots, from $z_{\rm p}=z_{\rm d}+\Delta z$ to $z_{\rm d}$, and obtain the mean number of mergers in each mass-ratio bin, $\langle \Delta N_{\rm merge}(\xi|M,z_{\rm d})\rangle$. 
Similarly, we measure the logarithmic mass growth of the main branch, $\Delta \log M=\log M_0-\log M_1$, and obtain the mean value $\langle \Delta \log M\rangle$. 
The specific merger rate is then estimated as $\langle \Delta N_{\rm merge}(\xi|M,z_{\rm d})\rangle/[\Delta \xi \langle \Delta \log M\rangle]$. 
By integrating this expression over $\xi$, we also compute the cumulative specific merger rate, $\langle N_{\rm merge}(>\xi_{\rm min}|M,z_{\rm d})\rangle/\langle \Delta \log M\rangle$. Following \citet{Fakhouri2010}, we also compute the mean merger rate per halo,
$dN_{\rm merger}/d\xi/dz/N_{\rm halo},$
This quantity is evaluated using the same merger events and snapshot intervals described above.

\begin{figure*}[!t]
    \centering
    \includegraphics[width=1\textwidth]{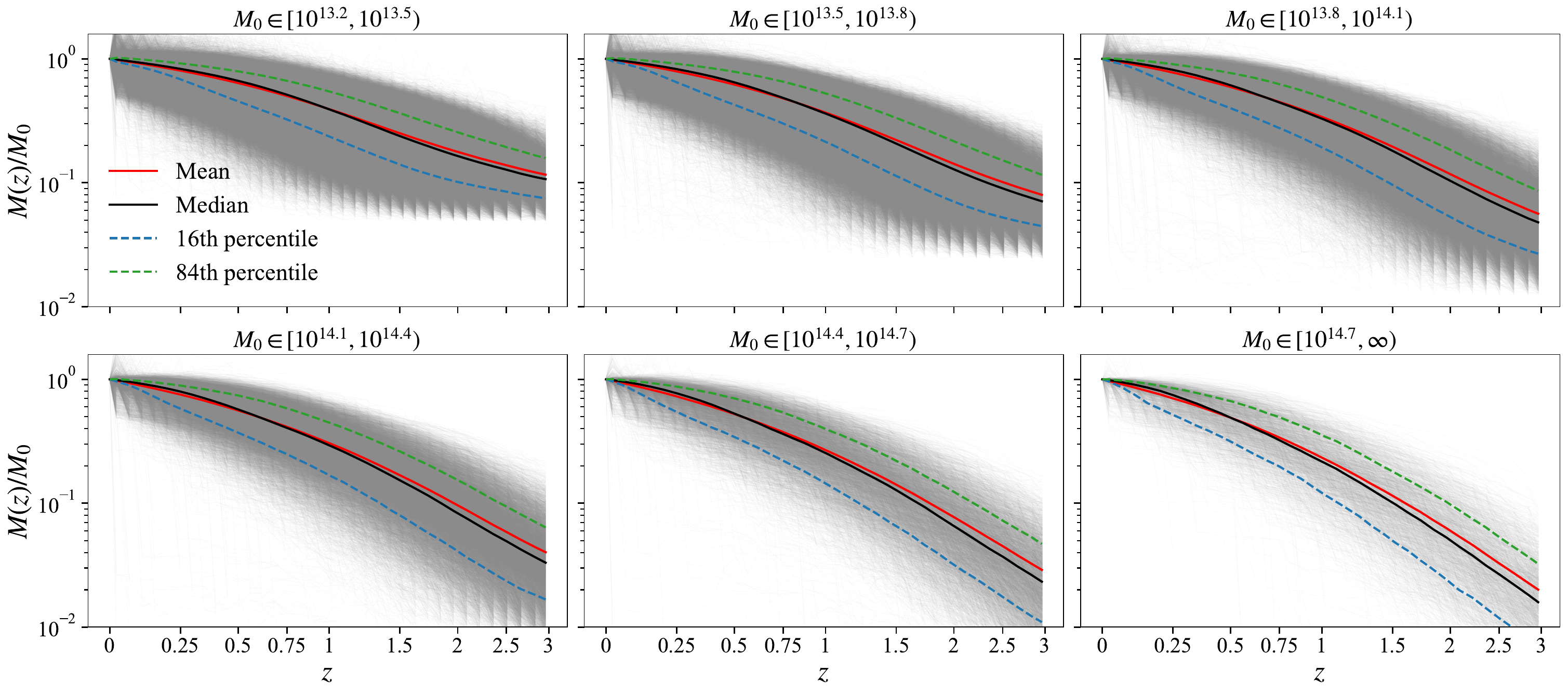}
    \caption{
    Mass accretion histories of individual halos in GR for different halo mass bins at $z=0$ .
    The halo mass evolution is constructed by tracing the main branch progenitor back in time and is normalized by the final halo mass, $M_0$, i.e., $M(z)/M_0$.
    Grey lines represent individual halos, while the red and black solid lines show the mean and median MAHs, respectively.
    The blue and green dashed lines indicate the 16th and 84th percentiles of the distribution.
    }
    \label{fig:mah_all}
\end{figure*}

\section{Results} \label{sec:results}

In this section, we present a comparative analysis of the mass accretion histories and merger rates of dark matter halos across different gravity models.

\subsection{Mass Accretion Histories} \label{subsec:mah}

Figure \ref{fig:mah_all} shows the normalized MAHs of individual halos in GR for different halo mass bins. $M_0$ denotes the halo mass at $z=0$.  
Each grey curve corresponds to one halo, while the red and black solid curves show the mean and median MAHs, respectively. The blue and green dashed curves denote the 16th and 84th percentiles, which characterize the scatter among individual halo mass growth histories. The vertical band-like structures in the grey curves arise because the MAHs are sampled at discrete simulation snapshots. In the low-mass bins, the lower envelope of the individual MAHs becomes nearly horizontal at high redshift. 
For example, in the mass bin $M_0 \in [10^{13.2}, 10^{13.5})\,h^{-1}{\rm M_\odot}$, this flattening is visible at $z \gtrsim 1$. This behavior is caused by the finite halo mass resolution. 

To minimize the numerical resolution effects, we retain only those data points for which fewer than 10\% of the halos in the sample (selected at z = 0) fall below the HBT resolution limit (20 particles) at the corresponding redshift. In other words, a data point is included in the mean MAHs shown in Fig. \ref{fig:mah_all} only if the fraction of halos in the sample that become unresolved by HBT at that redshift is below 10\%; otherwise, that data point is not shown.

\begin{figure*}[!t]
    \centering
    \includegraphics[width=1\textwidth]{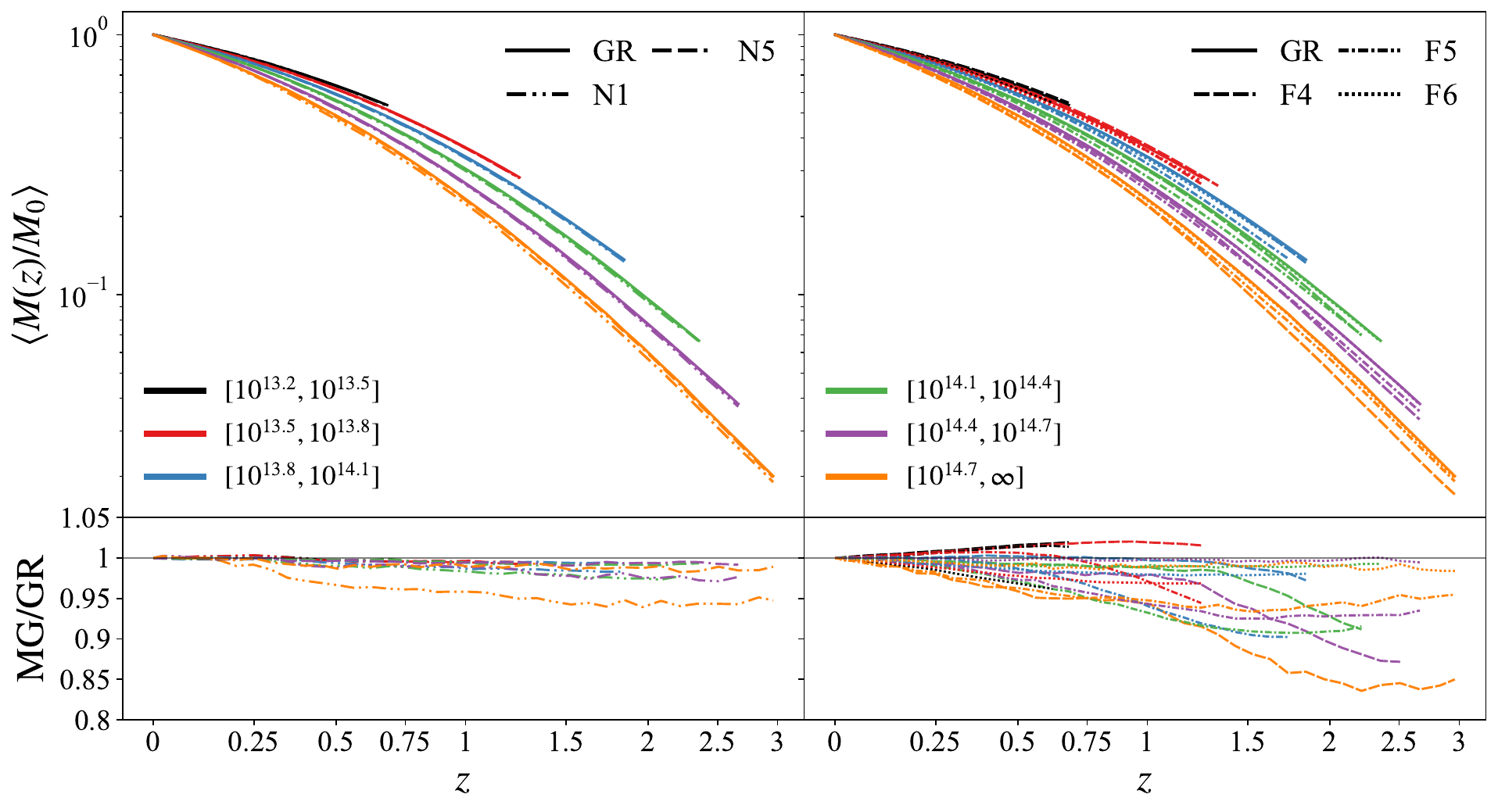}
    \caption{
    Comparison of the mean mass accretion histories of dark matter halos in different gravity models.
    The halo mass evolution is normalized by the final halo mass, $M_0$, and is shown as $M(z)/M_0$.
    Different colors denote different $z=0$ halo mass bins, while different line styles correspond to different gravity models.
    The left panels compare GR with the nDGP models, while the right panels compare GR with the $f(R)$ models.
    The upper panels show the mean normalized MAHs, and the lower panels show the ratio of each modified gravity model compared to GR.
    }
    \label{fig:mah_mean}
\end{figure*}

Figure \ref{fig:mah_mean} compares the mean normalized MAHs among different gravity models. 
Different colors denote different $z=0$ halo mass bins, while different line styles represent different gravity models. 
The upper panel shows the mean mass accretion histories, $\langle M(z)/M_0\rangle $, and the lower panel shows the ratio of each modified gravity model to GR. Note that halos selected within the same $z=0$ mass bin in different gravity models do not necessarily correspond to the same objects in the initial conditions, owing to differences in their mass accretion histories.

For the nDGP models, N1 shows a larger departure from GR than N5, consistent with its stronger modification of gravity. The deviation also increases with halo mass, reaching up to about $5\%$ for the most massive sample. This indicates that halos with the same present-day mass generally assembled more rapidly in nDGP than in GR at earlier times. The differences are considerably larger in the $f(R)$ models. In particular, the deviations can exceed $10\%$ in F5 and $15\%$ in F4 at high redshift. This demonstrates that the fifth force can substantially accelerate halo mass growth when the screening is inefficient. In addition, we note that for the lower-mass samples in F4 and F5, the ratio can slightly exceed unity at earlier times. This indicates that these halos assemble more rapidly than their GR counterparts during the early stages of their evolution, followed by relatively slower growth at later times. In general, the results indicate that modifications to GR can leave a noticeable imprint on the growth and evolution of dark matter halos, and the imprint can depend on the details of how the gravity model is changed.

\subsection{Halo Merger Rates} \label{subsec:merger_rate}

\begin{figure}[!tb]
    \centering
    \includegraphics[width=1\columnwidth]{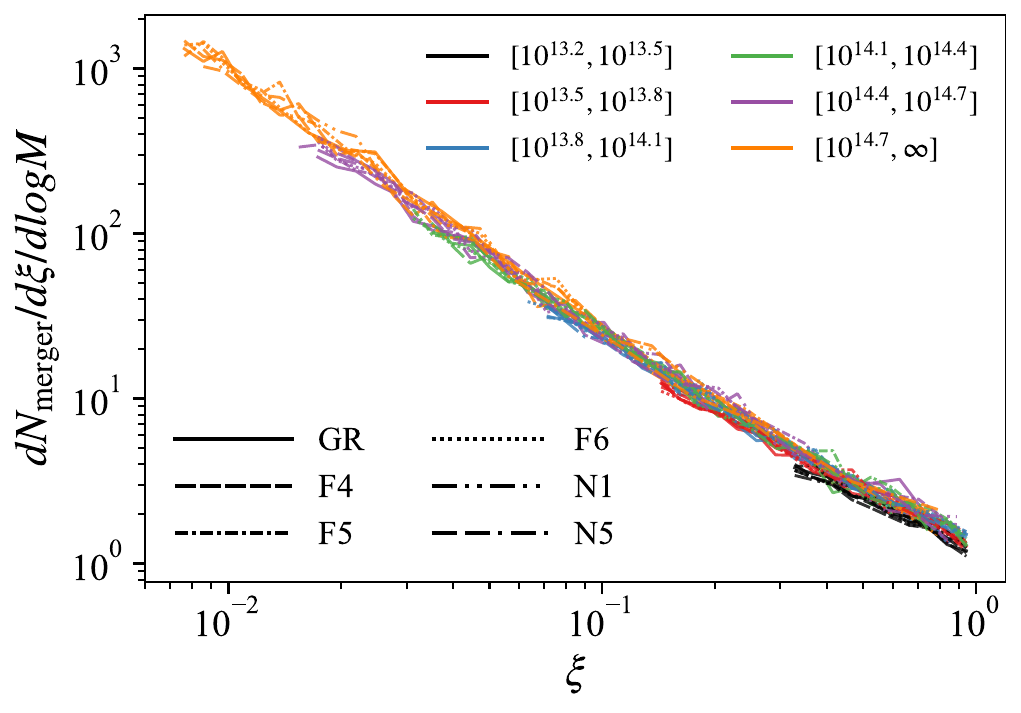}
    \caption{
    Specific merger rate of dark matter halos in different gravity models, measured over the redshift interval $(z_{\rm d},z_{\rm p})=(0.02 ,0.06)$.
    The horizontal axis shows the merger mass ratio. Different colors represent different halo mass bins, while different line styles correspond to different gravity models. The merger rates exhibit similar trends across the different gravity models.
    }
    \label{fig:specific_merger_rate_all_sims}
\end{figure}

\begin{figure}[!tb]
    \centering
    \includegraphics[width=1\columnwidth]{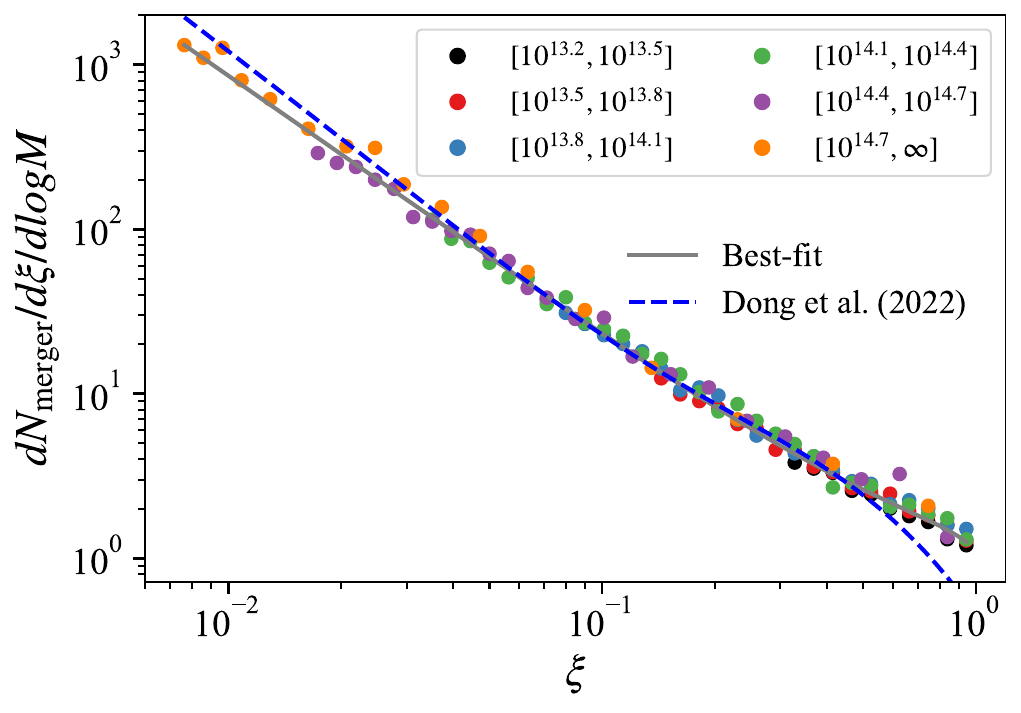}
    \caption{
    Best-fit result for the GR specific merger rate using the double-Schechter function in Eq. (\ref{eq:double_schechter_merger_rate}).
    The horizontal and vertical axes, as well as the colors of the data points, are the same as those in Fig. \ref{fig:specific_merger_rate_all_sims}.
    The thin grey curves show the best-fit model. The blue dashed curves show the fitting formula of \citet{Dong2022}.}
    \label{fig:fit_gr}
\end{figure}

Following the procedure described in Section \ref{subsec:merger_rates_mah}, we compute the specific merger rate of dark matter halos for different gravity models.
Figure \ref{fig:specific_merger_rate_all_sims} shows the specific merger rate as a function of the merger mass ratio, $\xi$, measured over the redshift interval $(z_{\rm d},z_{\rm p})=(0.02,0.06)$.
Different colors represent different halo mass bins, while different line styles correspond to different gravity models.

First of all, the specific merger in different gravity models exhibit similar trends as functions of $\xi$.
The results for GR, $f(R)$ gravity, and the nDGP models largely overlap over the range of merger mass ratios considered here, indicating that the halo merger rate preserves a high degree of universality across different gravity models. Second, within each individual gravity model, the dependence of the specific merger rate on halo mass is relatively weak.
Although the halos are divided into different mass bins, the corresponding curves remain close to one another and approximately follow the same functional form.

To further quantify this behavior, we fit the GR merger rate using a double-Schechter function \citep{Han2018, Dong2022},
\begin{equation}
dN_{\rm merger}/d\xi/d\log M = \left(a_1 \xi^{b_1} + a_2 \xi^{b_2}\right)
\exp\left(c\xi^d\right),
\label{eq:double_schechter_merger_rate}
\end{equation}
where $a_1$, $a_2$, $b_1$, $b_2$, $c$ and $d$ are free parameters. For the GR simulation, we obtain the best-fit parameter vector $(a_1,a_2,b_1,b_2,c,d)$=$(0.57,0.80,-1.58,-0.13,-0.20,6.62)$. The fitting result is shown in Fig. \ref{fig:fit_gr}. 
The colors have the same meaning as in Fig. \ref{fig:specific_merger_rate_all_sims}, while the thin grey curves show the best-fit model. 
The double-Schechter-like function provides a good description of the GR merger rate over the range of $\xi$ and halo mass considered in this work. Our measurements are also in reasonable agreement with the fitting result of \citet{Dong2022} but we note that near $\xi\simeq1$, our merger rate is somewhat higher. This difference mainly arises from the slightly different definitions of the merger mass ratio. In \citet{Dong2022}, the mass ratio is defined relative to the main progenitor recorded by the \textsc{HBT} subhalo finder, which is not necessarily the most massive progenitor and results in 
a fraction of merger events to have $\xi>1$. Our measurements near $\xi\simeq1$ effectively include the contribution that would appear at $\xi>1$ under their definition, thus degrading the overall agreement. Towards the low $\xi$ end, our measurements tend to be slightly lower than that of \citet{Dong2022}. This again may be attributed to systematics in the mass definitions. \citet{Jiang2025} adopted halo mass definitions that yields values comparable to ours, and also found a merger rate slightly lower than the prediction of \citet{Dong2022} for mass ratios above $10^{-3}$.

\begin{figure*}[!tb]
    \centering
    \includegraphics[width=1\textwidth]{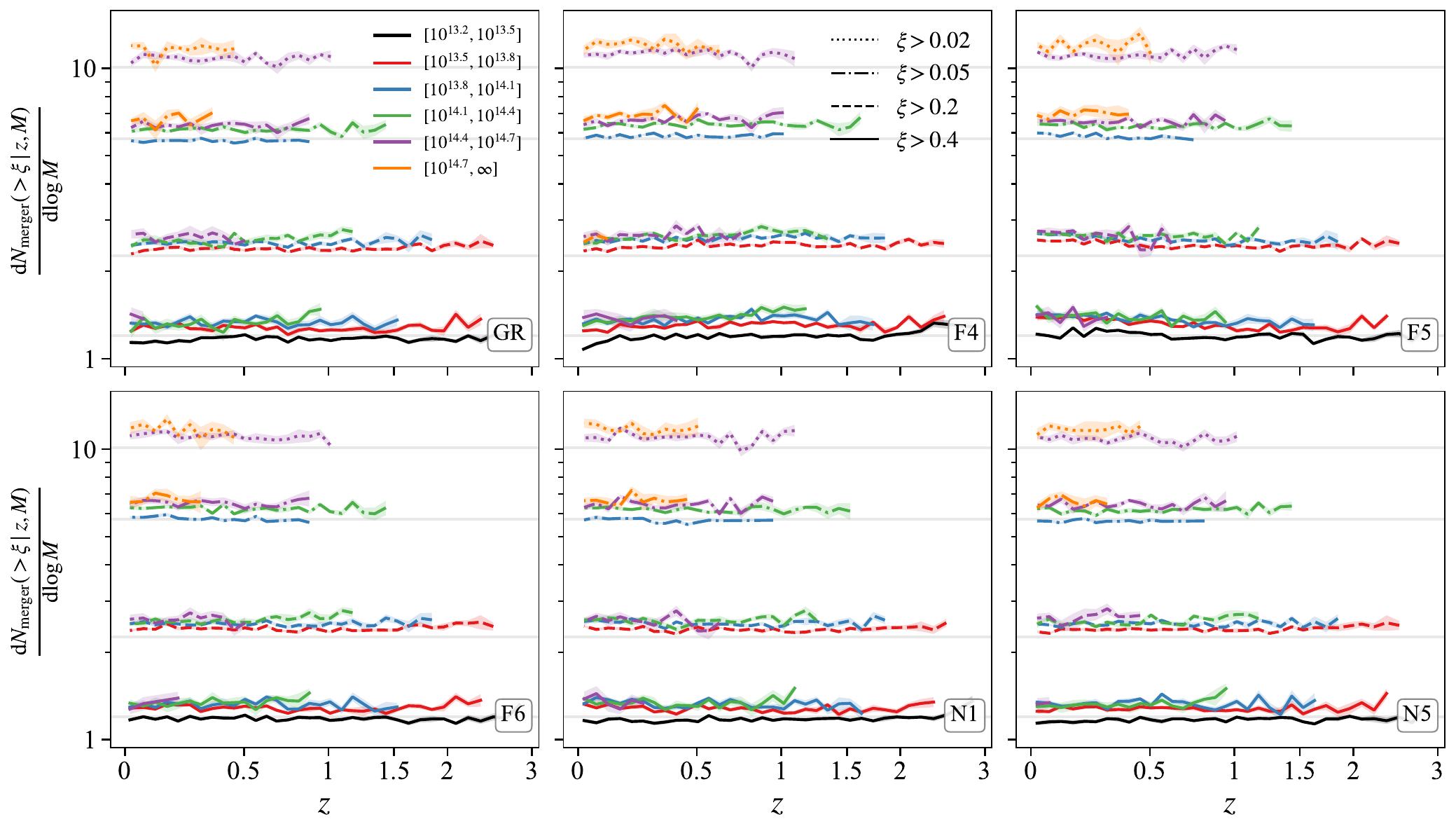}
    \caption{
    Cumulative specific merger rate, as a function of redshift for different gravity models. 
    Different colors denote different halo mass bins, while different line styles correspond to four $\xi_{\rm min}$ thresholds. The shaded regions indicate the $1\sigma$ statistical uncertainties, estimated for each curve using jackknife resampling.
    The light grey lines show the integrated best-fit GR merger rate from Figure \ref{fig:fit_gr}. 
    All six simulations exhibit nearly redshift-independent merger rates.
    }
    \label{fig:mr_dong}
\end{figure*}

Figure \ref{fig:mr_dong} shows the cumulative specific merger rate,
$dN_{\rm merger}(>\xi_{\rm min}|z,M)/d \log M$, for different gravity models as a function of redshift. 
Each panel corresponds to one gravity model, with the horizontal axis representing redshift and the vertical axis showing the cumulative merger count per unit logarithmic halo mass growth. 
We consider four thresholds for the merger mass ratio, $\xi_{\rm min}=\{0.02, 0.05, 0.2, 0.4\}$, indicated by different line styles, while different colors denote different halo mass bins. When generating Fig. \ref{fig:mr_dong}, we exclude the first redshift interval, $(z_{\rm d},z_{\rm p})=(0 ,0.02)$, because the spacing to the nearest simulation snapshot is $\Delta z \sim 0.04$, which can affect the measured merger rates \citep{Fakhouri2008}.

For lower-mass halos, the finite mass resolution limits the measurements at small $\xi_{\rm min}$. For higher-mass halos, high-mass-ratio mergers are intrinsically rare, so the cumulative counts at large $\xi_{\rm min}$ suffer from substantial statistical noise and we simply exclude them from the analysis. As a result, only the measurements with smaller $\xi_{\rm min}$ are shown for the highest-mass bins.
The light grey lines show the cumulative specific merger rate obtained by integrating the best-fit GR specific merger rate from Fig. \ref{fig:fit_gr} and Eq. (\ref{eq:double_schechter_merger_rate}).
The shaded regions indicate the $1\sigma$ uncertainties of the measurements, which are estimated using a jackknife resampling algorithm by dividing the simulation box into eight equal-sized subvolumes.

\begin{figure}[!tb]
    \centering
    \includegraphics[width=1\columnwidth]{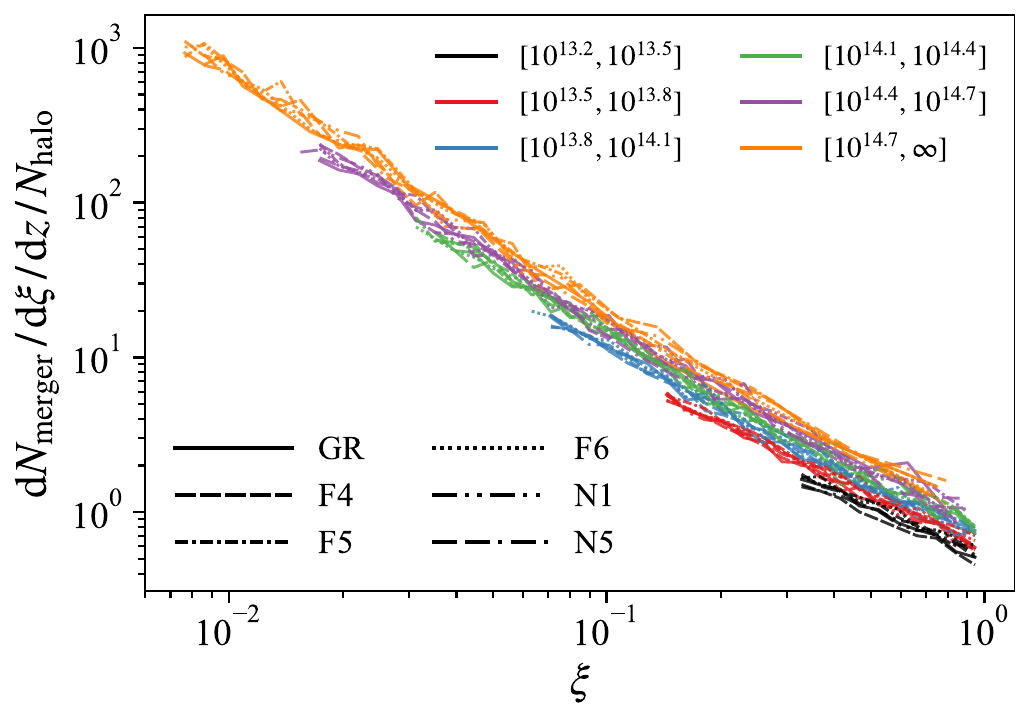}
    \caption{
    Mean merger rate per halo as a function of merger mass ratio 
    for different gravity models, measured over the redshift interval 
    $(z_{\rm d}, z_{\rm p}) = (0.02, 0.06)$.
    }
    \label{fig:mr3_mh_all}
\end{figure}

\begin{figure*}[!tb]
    \centering
    \includegraphics[width=1\textwidth]{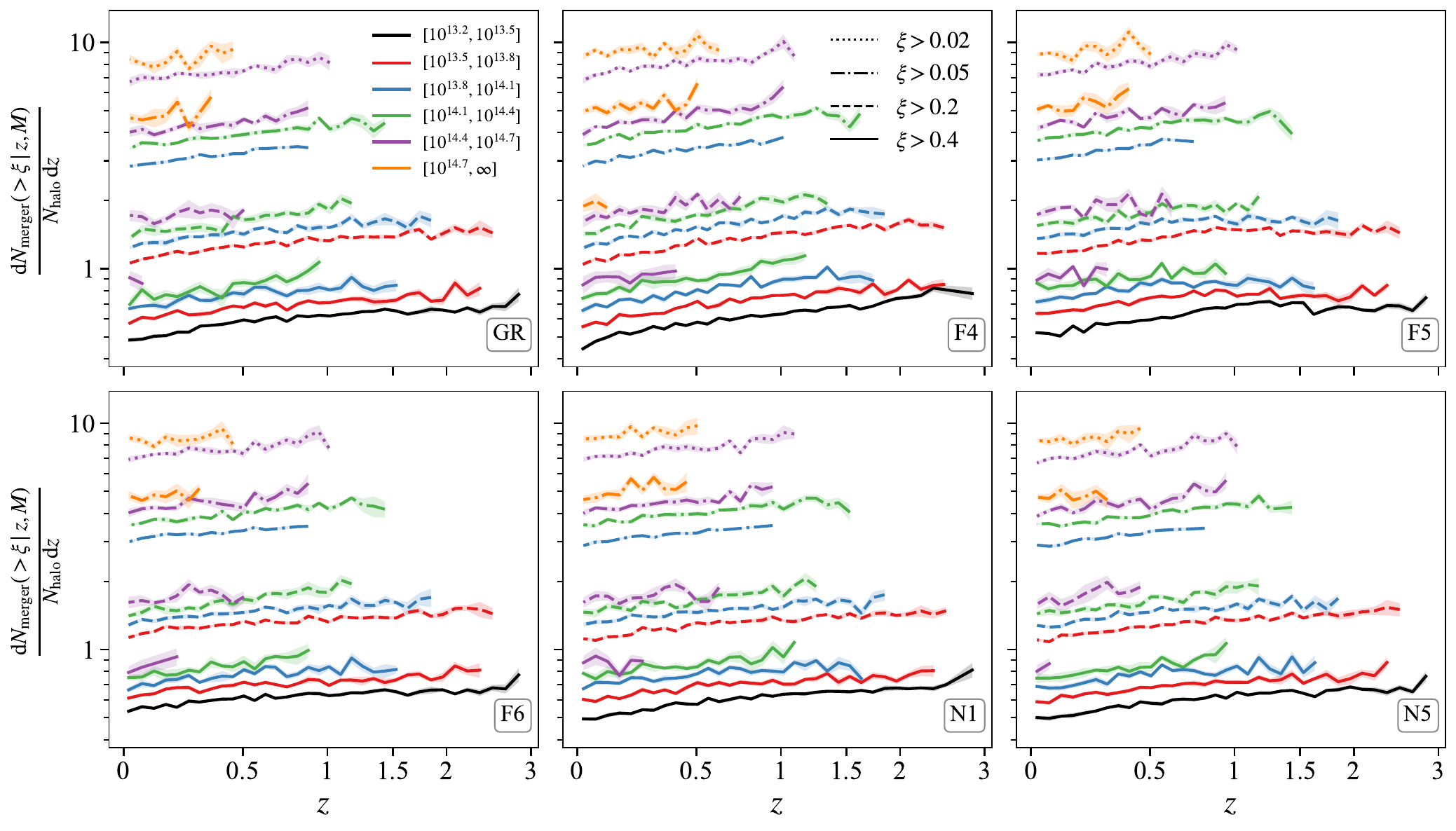}
    \caption{
    Redshift evolution of the mean halo merger rate per halo, $dN_{\rm merge}/dz/N_{\rm halo}$, for different gravity models.
    The colors denote different halo mass bins, while the line styles correspond to different thresholds in merger mass-ratio thresholds, $\xi_{\rm min}$, as in Fig. \ref{fig:mr_dong}.
    }
    \label{fig:mr_mh}
\end{figure*}
As expected, for all six simulations, the cumulative merger counts remain nearly independent of redshift, and the values are consistent across different $\xi_{\rm min}$ thresholds and halo mass bins. 
In addition, we note that the grey curves obtained by integrating the best-fit GR specific merger rate do not perfectly match the measured curves at all $\xi_{\rm min}$ thresholds. 
This discrepancy arises because the specific merger rate retains a residual dependence on halo mass. 
During the fitting process, lower-mass halos contribute more strongly to the fit because of their higher abundance. 
Consequently, the integrated grey curves tend to agree more closely with the lower mass bins that dominate the fit at each $\xi_{\rm min}$ threshold, while the measured specific merger rates for higher-mass halos generally exceed the fitted predictions.

We examine whether the universal functional form of the merger rate proposed by \citet{Fakhouri2010} is preserved in modified gravity simulations. Following their definition, we compute
$
    B(M_0,\xi)/n(M_0) \equiv 
    dN_{\rm merger}/d\xi/dz/N_{\rm halo} ,\label{eq:mergrerate_dz}
$
where $B(M_0,\xi)$ is the merger rate of descendant halos with mass $M_0$ and progenitor mass ratio $\xi$, and $n(M_0)$ is the number density of descendant halos in the corresponding mass bin. The result is shown in Fig. \ref{fig:mr3_mh_all}. At fixed halo mass, the mean merger rates per halo measured in different gravity models are broadly consistent, with no clear differences among the models. However, within a given gravity model, the merger rate exhibits a pronounced dependence on descendant halo mass. This is simply because the specific merger rate, 
Equation~\ref{eq:specific_merger_rate}, is expected to be universal~\citep{Dong2022,Jiang2025} following the unbiased accretion framework \citep{Han16}. The ``absolute'' merger rate defined in Equation~\eqref{eq:mergrerate_dz}, however, differs from it by a factor of the halo growth rate, $\mathrm{d}\log M/\mathrm{d}z$. As high mass halos form recently and have higher accretion rate on average (see Fig.~\ref{fig:mah_mean}), the absolute merger rate is thus expected to be higher in more massive halos.

We also present the mean merger rate per halo per unit redshift, 
$dN_{\rm merger}/dz/N_{\rm halo}$, in Fig. \ref{fig:mr_mh}.
The line styles and colors are the same as those in Fig. \ref{fig:mr_dong}, corresponding to different merger mass-ratio thresholds and halo mass bins, respectively. 
All gravity models show only a weak dependence on redshift over the range considered here, with no significant differences among the different gravity models. This is similar to our previous findings.

\begin{figure*}[!tb]
    \centering
    \includegraphics[width=1\textwidth]{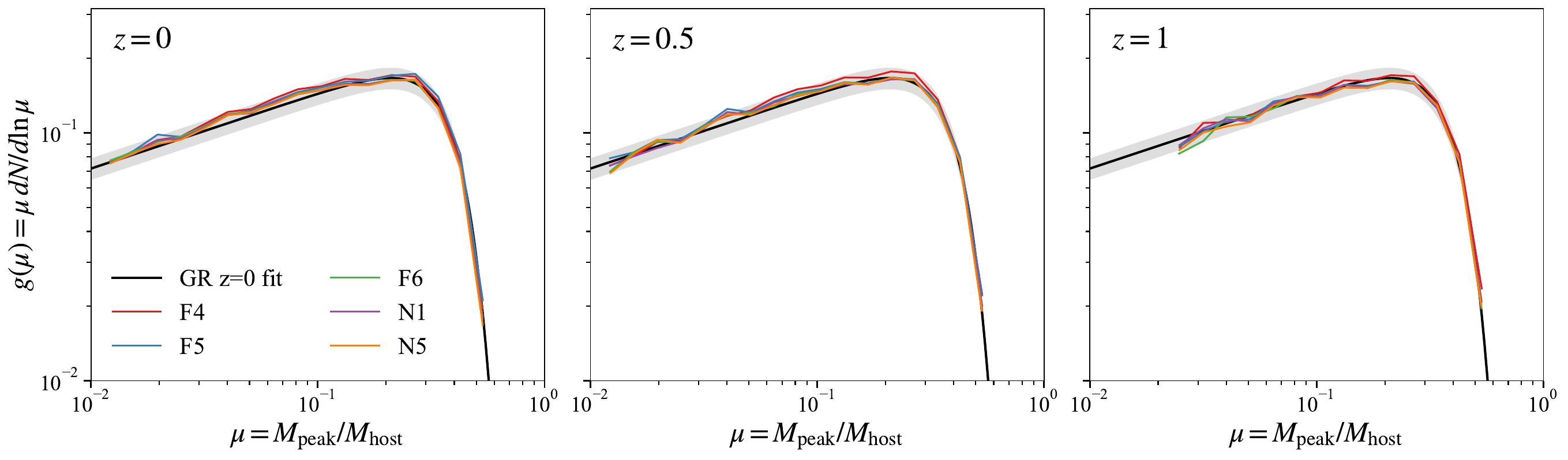}
    \caption{Level-1 peak mass functions (PMFs) measured at three redshifts for different gravity models. The black solid curve shows the double-Schechter fit to the GR measurement at redshift $z=0$, and the gray shaded region indicates the $\pm 10\%$ deviation from this fit. Colored curves show the measured PMFs for the other gravity models. The curves shown here are obtained by averaging the PMFs measured in different host halo mass bins above $10^{13.2}\,h^{-1}{\rm M_\odot}$, weighted by the number of subhaloes in each bin.}
    \label{fig:pmf}
\end{figure*}

On the other hand, the peak mass function (PMF) provides a description of the merging population by quantifying the accumulated number of accreted progenitor halos as a function of their peak-to-host mass ratio \citep{Jiang2025}. As shown by \citet{Dong2022,Jiang2025}, the specific merger rate is the derivative of the PMF.
We therefore also compare the PMFs measured in different gravity models, as shown in Fig.~\ref{fig:pmf}. In GR, the PMFs show good agreement across different host halo masses and redshifts \citep{Jiang2025}. We find that the merger rate exhibits only a very weak dependence on host halo mass in the modified gravity models. The PMFs shown in Fig.~\ref{fig:pmf} are therefore obtained by averaging the PMFs measured in different host halo mass bins above $10^{13.2}\,h^{-1}{\rm M_\odot}$, weighted by the number of subhaloes in each bin. The black solid curve shows the double-Schechter fit to the GR measurement at $z=0$, while the colored curves show the results of different modified gravity models at the three redshifts. We find that most of the models have a PMF in agreement with the GR result within $\pm 10\%$ region for $\mu < 0.6$ at all three redshifts, suggesting that the redshift independence of the PMF is largely preserved in these models.

\section{Discussion: the influence of halo mass definition}
\label{sec:discussion}

\begin{figure*}[!tb]
    \centering
    \includegraphics[width=1\textwidth]{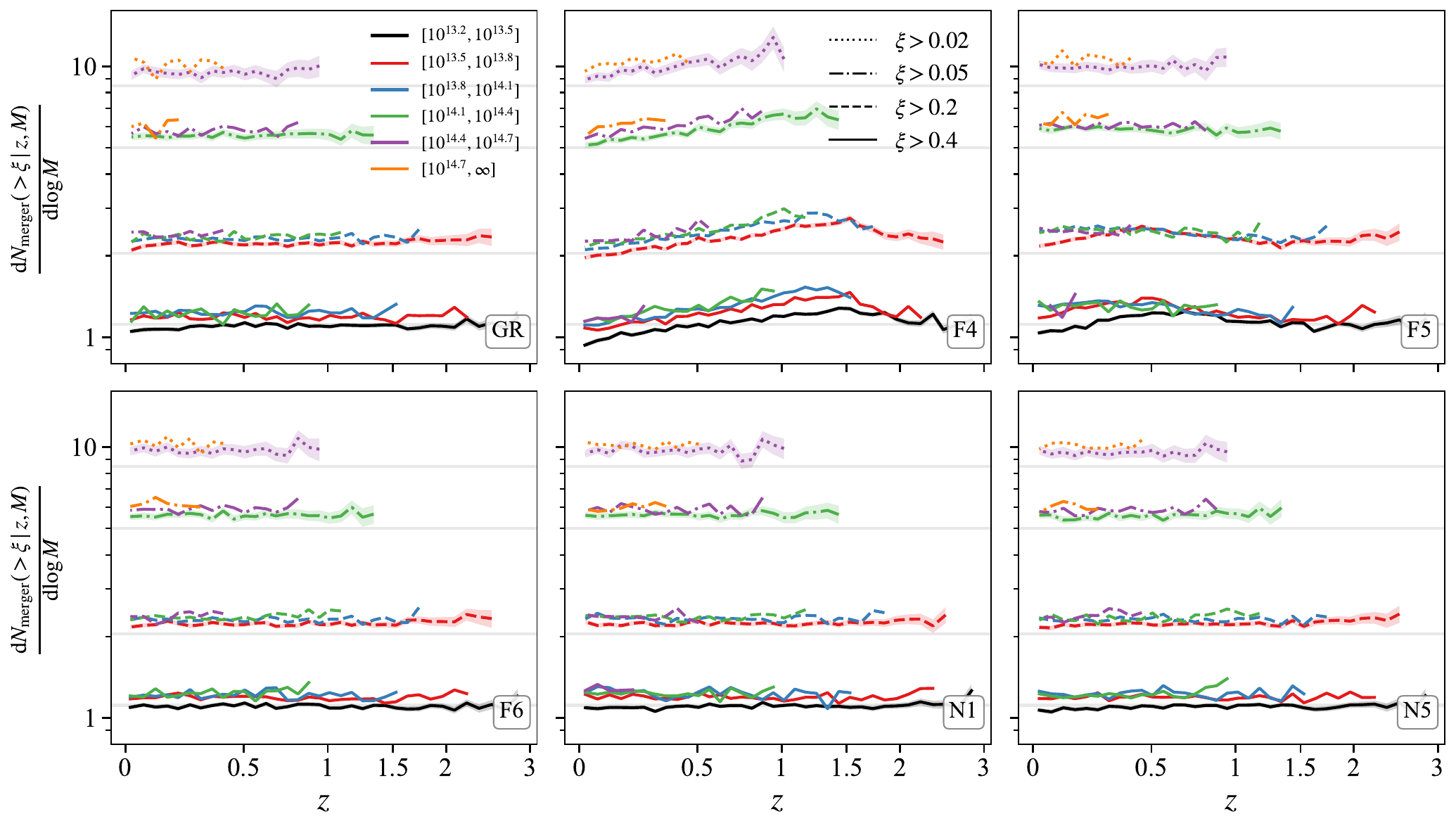}
    \caption{
    Same as Fig.~\ref{fig:mr_dong}, but using the halo mass defined as the sum of the bound masses of all subhalos within each FoF halo.
    With this mass definition, F4 and F5 exhibit more pronounced deviations from redshift independence, i.e. more violation of universality.
    }
    \label{fig:mr_dong_Bound}
\end{figure*}

\begin{figure}[!htb]
    \centering
    \includegraphics[width=1\columnwidth]{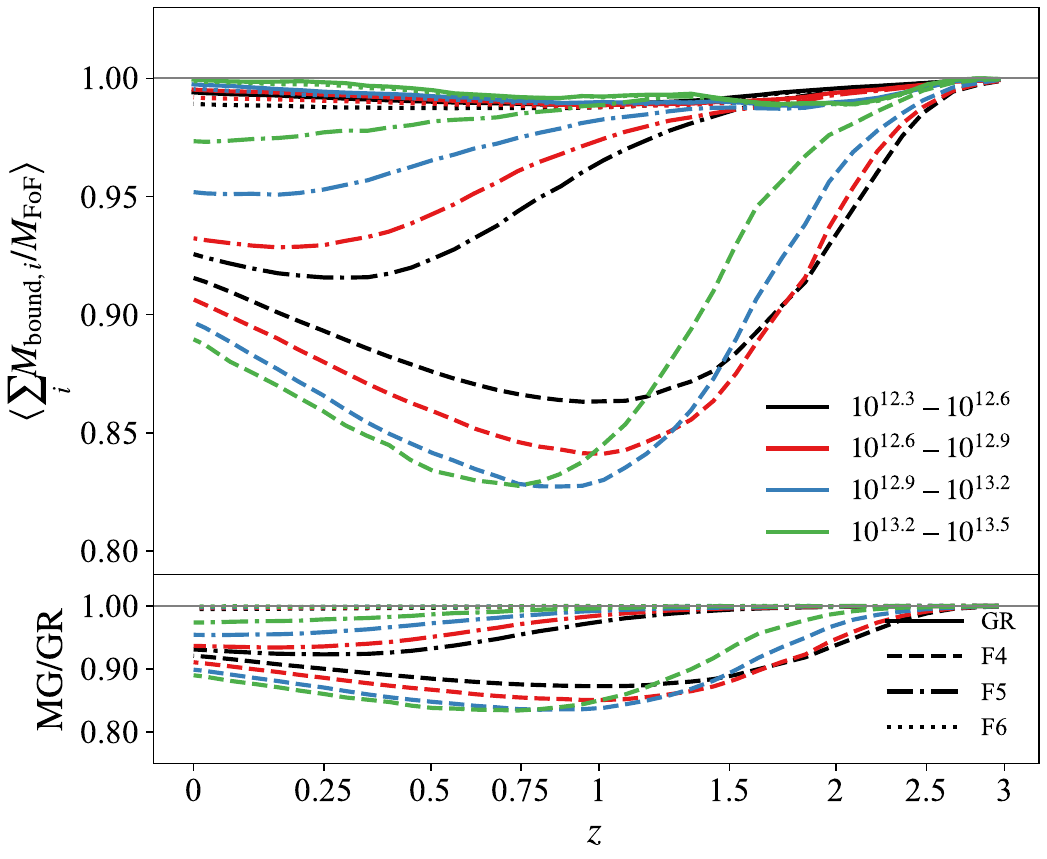}
    \caption{
Evolution of the mean ratio of the summation of the subhalo bound mass to the FoF halo mass in different gravity models.
Halos are matched across the GR and $f(R)$ simulations using particle IDs at $z\simeq3$ and divided into four mass bins according to their GR FoF masses at the matching redshift, as indicated by different colors.
The upper panels show the mean ratio of the sum of the bound masses of all subhalos to the FoF halo mass, with different line styles representing different gravity models.
The lower panels show the corresponding ratios of the $f(R)$ models to GR.
    }
    \label{fig:bf}
\end{figure}

Previous studies on the universality of the specific merger rate have adopted varying definitions of the halo mass. \citet{Dong2022} adopted the virial mass and applied some post-processing to the merger tree to account for backsplash and flyby halos to achieve universality. On the other hand, \citet{Jiang2025} worked with the original \textsc{HBT+} merger tree. They investigated different mass definitions and found that adopting the summation of the bound mass from all subhalos in each FoF halo as the host mass yields the highest universality. 

In our analysis, we also follow the original \textsc{HBT+} tree. To compare with the findings of \citet{Jiang2025}, 
we repeat the analysis following their halo mass definition. 
The result is shown in Fig.~\ref{fig:mr_dong_Bound}. While the merger-rate universality is well preserved in GR, F6, N1 and N5, significant deviations from redshift independence arise in F4 and F5. Since the underlying simulations and analysis procedure are otherwise unchanged, these differences can be directly attributed to the adopted halo mass definition.

The halo masses used here are bound masses returned by \textsc{HBT+}. During the unbinding step, the code evaluates whether a particle is bound using its kinetic energy and the Newtonian gravitational potential generated by the candidate structure \citep{Han2012, Han2018}. The fifth-force contribution to the binding potential is not included. This distinction is unimportant when the fifth force is screened, but it matters in an unscreened halo. The particle velocities in the simulation respond to the total force, whereas the unbinding calculation compares those velocities with a shallower, Newtonian-only potential. Some particles that are bound by the combined Newtonian and fifth forces can consequently be classified as unbound. Their mass is then omitted from the subhalo, and because the alternative host-halo mass is defined as the sum of the bound masses of its subhalos, the same bias propagates to the host halo mass.

To see the influence of unbinding directly, we compare the bound mass fraction between GR and $f(R)$ models in Fig.~\ref{fig:bf}. To make sure we compare the same set of halos across models and time, we first match halos across the GR and $f(R)$ simulations using particle IDs at a high initial redshift ($z\simeq3$) available to our analysis, where the difference among the simulations is minimal. We then divide the halos into four mass bins according to their GR FoF masses at the initial redshift. The matched halos are then tracked along their respective main branches, and the evolution of the mean ratio of the summation subhalo bound mass to the FoF mass is shown in Fig.~\ref{fig:bf}. Consistent with our expectations, F4 and F5 exhibit systematically lower bound-mass fractions than GR, with differences that vary with redshift and halo mass. These results show that the two mass definitions yield similar halo masses in GR and F6, allowing merger rate universality to be well preserved under either definition. However, the F4 and F5 models have much larger differences that lead to pronounced deviations from the universal behavior when the bound mass definition is adopted. A strong MG effect of f(R) gravity can produce a evolution history more complicated than GR that the mass fraction of the bound structures within a halo is sensitive to redshift.

The same mass-estimation bias can also affect the progenitor mass ratio. In a merger, the main progenitor is more massive than the secondary progenitor and can therefore be affected more strongly by the reduction in the measured bound mass. Its mass, which appears in the denominator of the merger ratio, is then underestimated by a larger fraction, shifting the measured value of $\xi$ upward. Put another way, a merger assigned to a given $\xi$ bin can correspond to a smaller-$\xi$ event when the mass bias is removed. Because the specific merger rate rises toward smaller merger ratios, moving these more numerous events into larger measured-$\xi$ bins can enhance the measured merger rate. This provides an important mechanism through which the mass-estimation bias can contribute to the apparent deviation from merger-rate universality in F4 and F5.

The different levels of deviation among the gravity models are broadly consistent with their screening mechanisms. In the nDGP models, the Vainshtein radius of a massive halo is much larger than the halo radius. For example, a halo with $R_{200}=1\,{\rm Mpc}$ at $z=0$ has a Vainshtein radius of approximately $40$ and $20\,{\rm Mpc}$ in N5 and N1, respectively. The fifth force is therefore efficiently suppressed on the scales relevant to halo binding, consistent with the close agreement of N1 and N5 with GR. In $f(R)$ gravity, the screening efficiency depends more sensitively on $f_{R0}$, halo mass, and environment \citep{Li2012a, Moran2015, Gronke2015}. F6 is largely screened in the massive halos considered here, F5 can be partially screened, and F4 remains unscreened over a much wider range of masses and redshifts. For a Virgo-like halo with $M_{200c}\sim10^{14}M_\odot$ and $r_{200c}\sim1h^{-1}{\rm Mpc}$, \citet{Moran2015} found transition redshifts of approximately $z\simeq4$, $1.5$, and $0$ for F4, F5, and F6, respectively. This ordering indicates that the transition to the unscreened regime occurs earlier in F4 and later in F5, consistent with the earlier onset of the merger rate deviation in F4 and the later deviation in F5. Consequently, neglecting the fifth force contribution in the unbinding calculation produces a larger and more redshift dependent mass bias in F4 and F5.

\section{Conclusions} \label{sec:conclusion}

In this work, we have investigated whether the universal behavior of the growth and merger histories of dark matter halos is preserved in modified gravity simulations.
Using the \textsc{ELEPHANT} suite, we studied halo mass accretion histories and merger rates in GR, three Hu–Sawicki $f(R)$ models, and two nDGP models.
Halo merger trees were constructed from the simulation outputs, allowing us to trace both the main branch mass growth and the merger events of dark matter halos.

For the mass accretion histories, modified gravity generally enhances halo growth relative to GR, reflecting the additional gravitational effects introduced by the fifth force. The enhancement is more pronounced in the stronger $f(R)$ models, with deviations from GR exceeding $10\%$ in F5 and $15\%$ in F4 over the redshift and mass ranges considered. This picture is in overall agreement with earlier studies that the modification of gravity can increase the halo mass function, i.e. the halos can form faster in MG models (\citealt{Gupta2022}). Therefore the halos in the MG simulations that have the same present day mass as GR preferentially have a lower mass progenitor along the growth or accretion history. 

We measured the halo merger rate using both the specific merger-rate definition of \citet{Dong2022} and the commonly adopted definition of \citet{Fakhouri2010}. Both measurements show broadly consistent behavior across all six gravity models, with the merger rate exhibiting little redshift dependence and preserving a nearly universal form. The \citet{Fakhouri2010} definition, however, retains a more pronounced dependence on halo mass because the mass dependence of the halo growth rate is not scaled out. In general, these results reveal the evidence of universality in the description of dark matter halos and the merger events, even though the underlying gravitational forces are changed by different levels. These results support the general ansatz that halos grow through approximately unbiased accretion~\citep{Han16,Dong2022,Jiang2025}, such that surrounding halos and diffuse matter are accreted indiscriminatively and in proportion to each other, irrespective of the exact strength or law of gravity. 

Finally, we examined the impact of the halo mass definition on the merger rate measurements. When the halo mass is instead defined as the sum of the bound masses of all subhalos, we can see a clear redshift-dependent deviations in F4 and F5, while the universality remains well preserved in the other models. We attribute these deviations primarily to the GR-only unbinding procedure in \textsc{HBT+}, which neglects the fifth-force contribution to the binding potential and can therefore underestimate subhalo masses, particularly when the fifth force is less efficiently screened. This mass-estimation bias can alter both the selected halo population and the measured progenitor mass ratios, thereby propagating into the merger rate statistics. Therefore, this violation can be understood as caused by the inconsistency of the mass definition adopted by the subhalo finder with the gravity model, instead of revealing any physical mechanism. Nevertheless, it highlights the crucial role of modifying subhalo finders in accordance with the modification of gravity, in order to identify generic physical laws governing structure formation in these general gravity models.

Because of the finite mass and temporal resolutions, our tests of halo-growth universality are restricted to massive halos and a relatively narrow redshift range.
Future work using larger-volume and higher-resolution simulations would allow this universality to be tested over a wider dynamic range in halo mass, redshift, and merger mass ratio.
In the present analysis, we adopt FoF masses to avoid the mass bias associated with the Newtonian only unbinding procedure. However, FoF groups are identified purely through configuration space linking and do not explicitly require all member particles to be gravitationally bound. A more physically motivated approach would be to incorporate the fifth force contribution into the unbinding procedure of \textsc{HBT+}, allowing bound halo and subhalo masses to be determined consistently with the modified gravity dynamics. Repeating the halo growth and merger rate measurements with such modified gravity consistent bound masses would provide an important test of the universality found here. 
Extensions to other dark matter and gravity models would further clarify the conditions under which the approximate universality of halo growth persists or breaks down, and may provide a more detailed connection between dark matter halos and galaxies.

\begin{acknowledgments}

HX thanks Wenkang Jiang for useful discussions. This work is supported by 
NSFC (12373003, 12595312), the National Key R\&D Program of China (2023YFA1605600, 2023YFA1607800, 2023YFA1607801), the China Manned Space Program (grant No. CMS-CSST-2025-A04), the Office of Science and Technology, Shanghai Municipal Government (grant Nos. 24DX1400100 and ZJ2023-ZD-001), and the 111 Project (No. B20019). HX and ZZ acknowledge the generous sponsorship from the Yangyang Development Fund. The computations in this paper were run on the Gravity Supercomputer at the Department of Astronomy, Shanghai Jiao Tong University. BL is supported by the European Research Council (ERC) Advanced Grant `UNCA', under the UKRI's Frontiers Research Guarantee No.~EP/Z533877/1, and the UK STFC Consolidated Grant No.~ST/X001075/1, and acknowledges the hosts by the National Astronomical Observatory of China in 2025/2026, during which part of this work was carried out. This work used the DiRAC@Durham facility managed by the Institute for Computational Cosmology on behalf of the STFC DiRAC HPC Facility (https://www.dirac.ac.uk). The equipment was funded by BEIS capital funding via STFC capital grants ST/K00042X/1, ST/P002293/1, ST/R002371/1 and ST/S002502/1, Durham University and STFC operations grant ST/R000832/1. DiRAC is part of the National e-Infrastructure.

\end{acknowledgments}

\bibliography{sample701,galaxysurvey}{}

@ARTICLE{AbdulKarim2025,
       author = {{Abdul Karim}, M. and {Aguilar}, J. and {Ahlen}, S. and {Alam}, S. and {Allen}, L. and {Allende Prieto}, C. and {Alves}, O. and {Anand}, A. and {Andrade}, U. and {Armengaud}, E. and {Aviles}, A. and {Bailey}, S. and {Baltay}, C. and {Bansal}, P. and {Bault}, A. and {Behera}, J. and {BenZvi}, S. and {Bianchi}, D. and {Blake}, C. and {Brieden}, S. and {Brodzeller}, A. and {Brooks}, D. and {Buckley-Geer}, E. and {Burtin}, E. and {Calderon}, R. and {Canning}, R. and {Rosell}, A. Carnero and {Carrilho}, P. and {Casas}, L. and {Castander}, F.~J. and {Charles}, M. and {Chaussidon}, E. and {Chaves-Montero}, J. and {Chebat}, D. and {Chen}, X. and {Claybaugh}, T. and {Cole}, S. and {Cooper}, A.~P. and {Cuceu}, A. and {Dawson}, K.~S. and {de la Macorra}, A. and {de Mattia}, A. and {Deiosso}, N. and {Della Costa}, J. and {Demina}, R. and {Dey}, A. and {Dey}, B. and {Ding}, Z. and {Doel}, P. and {Edelstein}, J. and {Eisenstein}, D.~J. and {Elbers}, W. and {Fagrelius}, P. and {Fanning}, K. and {Fern{\'a}ndez-Garc{\'\i}a}, E. and {Ferraro}, S. and {Font-Ribera}, A. and {Forero-Romero}, J.~E. and {Frenk}, C.~S. and {Garcia-Quintero}, C. and {Garrison}, L.~H. and {Gazta{\~n}aga}, E. and {Gil-Mar{\'\i}n}, H. and {Gontcho A Gontcho}, S. and {Gonzalez}, D. and {Gonzalez-Morales}, A.~X. and {Gordon}, C. and {Green}, D. and {Gutierrez}, G. and {Guy}, J. and {Hadzhiyska}, B. and {Hahn}, C. and {He}, S. and {Herbold}, M. and {Herrera-Alcantar}, H.~K. and {Ho}, M.-F. and {Honscheid}, K. and {Howlett}, C. and {Huterer}, D. and {Ishak}, M. and {Juneau}, S. and {Kamble}, N.~V. and {Kara{\c{c}}ayl{\i}}, N.~G. and {Kehoe}, R. and {Kent}, S. and {Kim}, A.~G. and {Kirkby}, D. and {Kisner}, T. and {Koposov}, S.~E. and {Kremin}, A. and {Krolewski}, A. and {Lahav}, O. and {Lamman}, C. and {Landriau}, M. and {Lang}, D. and {Lasker}, J. and {Le Goff}, J.~M. and {Le Guillou}, L. and {Leauthaud}, A. and {Levi}, M.~E. and {Li}, Q. and {Li}, T.~S. and {Lodha}, K. and {Lokken}, M. and {Lozano-Rodr{\'\i}guez}, F. and {Magneville}, C. and {Manera}, M. and {Martini}, P. and {Matthewson}, W.~L. and {Meisner}, A. and {Mena-Fern{\'a}ndez}, J. and {Menegas}, A. and {Mergulh{\~a}o}, T. and {Miquel}, R. and {Moustakas}, J. and {Mu{\~n}oz-Guti{\'e}rrez}, A. and {Mu{\~n}oz-Santos}, D. and {Myers}, A.~D. and {Nadathur}, S. and {Naidoo}, K. and {Napolitano}, L. and {Newman}, J.~A. and {Niz}, G. and {Noriega}, H.~E. and {Paillas}, E. and {Palanque-Delabrouille}, N. and {Pan}, J. and {Peacock}, J.~A. and {Pellejero Ibanez}, M. and {Percival}, W.~J. and {P{\'e}rez-Fern{\'a}ndez}, A. and {P{\'e}rez-R{\`a}fols}, I. and {Pieri}, M.~M. and {Poppett}, C. and {Prada}, F. and {Rabinowitz}, D. and {Raichoor}, A. and {Ram{\'\i}rez-P{\'e}rez}, C. and {Rashkovetskyi}, M. and {Ravoux}, C. and {Rich}, J. and {Rocher}, A. and {Rockosi}, C. and {Rohlf}, J. and {Rom{\'a}n-Herrera}, J.~O. and {Ross}, A.~J. and {Rossi}, G. and {Ruggeri}, R. and {Ruhlmann-Kleider}, V. and {Samushia}, L. and {Sanchez}, E. and {Sanders}, N. and {Schlegel}, D. and {Schubnell}, M. and {Seo}, H. and {Shafieloo}, A. and {Sharples}, R. and {Silber}, J. and {Sinigaglia}, F. and {Sprayberry}, D. and {Tan}, T. and {Tarl{\'e}}, G. and {Taylor}, P. and {Turner}, W. and {Ure{\~n}a-L{\'o}pez}, L.~A. and {Vaisakh}, R. and {Valdes}, F. and {Valogiannis}, G. and {Vargas-Maga{\~n}a}, M. and {Verde}, L. and {Walther}, M. and {Weaver}, B.~A. and {Weinberg}, D.~H. and {White}, M. and {Wolfson}, M. and {Y{\`e}che}, C. and {Yu}, J. and {Zaborowski}, E.~A. and {Zarrouk}, P. and {Zhai}, Z. and {Zhang}, H. and {Zhao}, C. and {Zhao}, G.~B. and {Zhou}, R. and {Zou}, H. and {DESI Collaboration}},
        title = "{DESI DR2 results. II. Measurements of baryon acoustic oscillations and cosmological constraints}",
      journal = {\prd},
         year = 2025,
        month = oct,
       volume = {112},
       number = {8},
          eid = {083515},
        pages = {083515},
          doi = {10.1103/tr6y-kpc6},
archivePrefix = {arXiv},
       eprint = {2503.14738},
 primaryClass = {astro-ph.CO},
       adsurl = {https://ui.adsabs.harvard.edu/abs/2025PhRvD.112h3515A}
}

@ARTICLE{Adame2025,
       author = {{Adame}, A.~G. and {Aguilar}, J. and {Ahlen}, S. and {Alam}, S. and {Alexander}, D.~M. and {Alvarez}, M. and {Alves}, O. and {Anand}, A. and {Andrade}, U. and {Armengaud}, E. and {Avila}, S. and {Aviles}, A. and {Awan}, H. and {Bahr-Kalus}, B. and {Bailey}, S. and {Baltay}, C. and {Bault}, A. and {Behera}, J. and {BenZvi}, S. and {Bera}, A. and {Beutler}, F. and {Bianchi}, D. and {Blake}, C. and {Blum}, R. and {Brieden}, S. and {Brodzeller}, A. and {Brooks}, D. and {Buckley-Geer}, E. and {Burtin}, E. and {Calderon}, R. and {Canning}, R. and {Carnero Rosell}, A. and {Cereskaite}, R. and {Cervantes-Cota}, J.~L. and {Chabanier}, S. and {Chaussidon}, E. and {Chaves-Montero}, J. and {Chen}, S. and {Chen}, X. and {Claybaugh}, T. and {Cole}, S. and {Cuceu}, A. and {Davis}, T.~M. and {Dawson}, K. and {de la Macorra}, A. and {de Mattia}, A. and {Deiosso}, N. and {Dey}, A. and {Dey}, B. and {Ding}, Z. and {Doel}, P. and {Edelstein}, J. and {Eftekharzadeh}, S. and {Eisenstein}, D.~J. and {Elliott}, A. and {Fagrelius}, P. and {Fanning}, K. and {Ferraro}, S. and {Ereza}, J. and {Findlay}, N. and {Flaugher}, B. and {Font-Ribera}, A. and {Forero-S{\'a}nchez}, D. and {Forero-Romero}, J.~E. and {Frenk}, C.~S. and {Garcia-Quintero}, C. and {Gazta{\~n}aga}, E. and {Gil-Mar{\'\i}n}, H. and {Gontcho a Gontcho}, S. and {Gonzalez-Morales}, A.~X. and {Gonzalez-Perez}, V. and {Gordon}, C. and {Green}, D. and {Gruen}, D. and {Gsponer}, R. and {Gutierrez}, G. and {Guy}, J. and {Hadzhiyska}, B. and {Hahn}, C. and {Hanif}, M.~M.~S. and {Herrera-Alcantar}, H.~K. and {Honscheid}, K. and {Howlett}, C. and {Huterer}, D. and {Ir{\v{s}}i{\v{c}}}, V. and {Ishak}, M. and {Juneau}, S. and {Kara{\c{c}}ayl{\i}}, N.~G. and {Kehoe}, R. and {Kent}, S. and {Kirkby}, D. and {Kremin}, A. and {Krolewski}, A. and {Lai}, Y. and {Lan}, T.-W. and {Landriau}, M. and {Lang}, D. and {Lasker}, J. and {Le Goff}, J.~M. and {Le Guillou}, L. and {Leauthaud}, A. and {Levi}, M.~E. and {Li}, T.~S. and {Linder}, E. and {Lodha}, K. and {Magneville}, C. and {Manera}, M. and {Margala}, D. and {Martini}, P. and {Maus}, M. and {McDonald}, P. and {Medina-Varela}, L. and {Meisner}, A. and {Mena-Fern{\'a}ndez}, J. and {Miquel}, R. and {Moon}, J. and {Moore}, S. and {Moustakas}, J. and {Mueller}, E. and {Mu{\~n}oz-Guti{\'e}rrez}, A. and {Myers}, A.~D. and {Nadathur}, S. and {Napolitano}, L. and {Neveux}, R. and {Newman}, J.~A. and {Nguyen}, N.~M. and {Nie}, J. and {Niz}, G. and {Noriega}, H.~E. and {Padmanabhan}, N. and {Paillas}, E. and {Palanque-Delabrouille}, N. and {Pan}, J. and {Penmetsa}, S. and {Percival}, W.~J. and {Pieri}, M.~M. and {Pinon}, M. and {Poppett}, C. and {Porredon}, A. and {Prada}, F. and {P{\'e}rez-Fern{\'a}ndez}, A. and {P{\'e}rez-R{\`a}fols}, I. and {Rabinowitz}, D. and {Raichoor}, A. and {Ram{\'\i}rez-P{\'e}rez}, C. and {Ramirez-Solano}, S. and {Rashkovetskyi}, M. and {Ravoux}, C. and {Rezaie}, M. and {Rich}, J. and {Rocher}, A. and {Rockosi}, C. and {Roe}, N.~A. and {Rosado-Marin}, A. and {Ross}, A.~J. and {Rossi}, G. and {Ruggeri}, R. and {Ruhlmann-Kleider}, V. and {Samushia}, L. and {Sanchez}, E. and {Saulder}, C. and {Schlafly}, E.~F. and {Schlegel}, D. and {Schubnell}, M. and {Seo}, H. and {Shafieloo}, A. and {Sharples}, R. and {Silber}, J. and {Slosar}, A. and {Smith}, A. and {Sprayberry}, D. and {Tan}, T. and {Tarl{\'e}}, G. and {Taylor}, P. and {Trusov}, S. and {Ure{\~n}a-L{\'o}pez}, L.~A. and {Vaisakh}, R. and {Valcin}, D. and {Valdes}, F. and {Vargas-Maga{\~n}a}, M. and {Verde}, L. and {Walther}, M. and {Wang}, B. and {Wang}, M.~S. and {Weaver}, B.~A. and {Weaverdyck}, N. and {Wechsler}, R.~H. and {Weinberg}, D.~H. and {White}, M. and {Yu}, J. and {Yu}, Y. and {Yuan}, S. and {Y{\`e}che}, C. and {Zaborowski}, E.~A. and {Zarrouk}, P. and {Zhang}, H. and {Zhao}, C. and {Zhao}, R. and {Zhou}, R. and {Zhuang}, T.},
        title = "{DESI 2024 VI: cosmological constraints from the measurements of baryon acoustic oscillations}",
      journal = {\jcap},
         year = 2025,
        month = feb,
       volume = {2025},
       number = {2},
          eid = {021},
        pages = {021},
          doi = {10.1088/1475-7516/2025/02/021},
archivePrefix = {arXiv},
       eprint = {2404.03002},
 primaryClass = {astro-ph.CO},
       adsurl = {https://ui.adsabs.harvard.edu/abs/2025JCAP...02..021A}
}

@ARTICLE{Dawson2016,
       author = {{Dawson}, Kyle S. and {Kneib}, Jean-Paul and {Percival}, Will J. and {Alam}, Shadab and {Albareti}, Franco D. and {Anderson}, Scott F. and {Armengaud}, Eric and {Aubourg}, {\'E}ric and {Bailey}, Stephen and {Bautista}, Julian E. and {Berlind}, Andreas A. and {Bershady}, Matthew A. and {Beutler}, Florian and {Bizyaev}, Dmitry and {Blanton}, Michael R. and {Blomqvist}, Michael and {Bolton}, Adam S. and {Bovy}, Jo and {Brandt}, W.~N. and {Brinkmann}, Jon and {Brownstein}, Joel R. and {Burtin}, Etienne and {Busca}, N.~G. and {Cai}, Zheng and {Chuang}, Chia-Hsun and {Clerc}, Nicolas and {Comparat}, Johan and {Cope}, Frances and {Croft}, Rupert A.~C. and {Cruz-Gonzalez}, Irene and {da Costa}, Luiz N. and {Cousinou}, Marie-Claude and {Darling}, Jeremy and {de la Macorra}, Axel and {de la Torre}, Sylvain and {Delubac}, Timoth{\'e}e and {du Mas des Bourboux}, H{\'e}lion and {Dwelly}, Tom and {Ealet}, Anne and {Eisenstein}, Daniel J. and {Eracleous}, Michael and {Escoffier}, S. and {Fan}, Xiaohui and {Finoguenov}, Alexis and {Font-Ribera}, Andreu and {Frinchaboy}, Peter and {Gaulme}, Patrick and {Georgakakis}, Antonis and {Green}, Paul and {Guo}, Hong and {Guy}, Julien and {Ho}, Shirley and {Holder}, Diana and {Huehnerhoff}, Joe and {Hutchinson}, Timothy and {Jing}, Yipeng and {Jullo}, Eric and {Kamble}, Vikrant and {Kinemuchi}, Karen and {Kirkby}, David and {Kitaura}, Francisco-Shu and {Klaene}, Mark A. and {Laher}, Russ R. and {Lang}, Dustin and {Laurent}, Pierre and {Le Goff}, Jean-Marc and {Li}, Cheng and {Liang}, Yu and {Lima}, Marcos and {Lin}, Qiufan and {Lin}, Weipeng and {Lin}, Yen-Ting and {Long}, Daniel C. and {Lundgren}, Britt and {MacDonald}, Nicholas and {Geimba Maia}, Marcio Antonio and {Malanushenko}, Elena and {Malanushenko}, Viktor and {Mariappan}, Vivek and {McBride}, Cameron K. and {McGreer}, Ian D. and {M{\'e}nard}, Brice and {Merloni}, Andrea and {Meza}, Andres and {Montero-Dorta}, Antonio D. and {Muna}, Demitri and {Myers}, Adam D. and {Nandra}, Kirpal and {Naugle}, Tracy and {Newman}, Jeffrey A. and {Noterdaeme}, Pasquier and {Nugent}, Peter and {Ogando}, Ricardo and {Olmstead}, Matthew D. and {Oravetz}, Audrey and {Oravetz}, Daniel J. and {Padmanabhan}, Nikhil and {Palanque-Delabrouille}, Nathalie and {Pan}, Kaike and {Parejko}, John K. and {P{\^a}ris}, Isabelle and {Peacock}, John A. and {Petitjean}, Patrick and {Pieri}, Matthew M. and {Pisani}, Alice and {Prada}, Francisco and {Prakash}, Abhishek and {Raichoor}, Anand and {Reid}, Beth and {Rich}, James and {Ridl}, Jethro and {Rodriguez-Torres}, Sergio and {Carnero Rosell}, Aurelio and {Ross}, Ashley J. and {Rossi}, Graziano and {Ruan}, John and {Salvato}, Mara and {Sayres}, Conor and {Schneider}, Donald P. and {Schlegel}, David J. and {Seljak}, Uros and {Seo}, Hee-Jong and {Sesar}, Branimir and {Shandera}, Sarah and {Shu}, Yiping and {Slosar}, An{\v{z}}e and {Sobreira}, Flavia and {Streblyanska}, Alina and {Suzuki}, Nao and {Taylor}, Donna and {Tao}, Charling and {Tinker}, Jeremy L. and {Tojeiro}, Rita and {Vargas-Maga{\~n}a}, Mariana and {Wang}, Yuting and {Weaver}, Benjamin A. and {Weinberg}, David H. and {White}, Martin and {Wood-Vasey}, W.~M. and {Yeche}, Christophe and {Zhai}, Zhongxu and {Zhao}, Cheng and {Zhao}, Gong-bo and {Zheng}, Zheng and {Ben Zhu}, Guangtun and {Zou}, Hu},
        title = "{The SDSS-IV Extended Baryon Oscillation Spectroscopic Survey: Overview and Early Data}",
      journal = {\aj},
         year = 2016,
        month = feb,
       volume = {151},
       number = {2},
          eid = {44},
        pages = {44},
          doi = {10.3847/0004-6256/151/2/44},
archivePrefix = {arXiv},
       eprint = {1508.04473},
 primaryClass = {astro-ph.CO},
       adsurl = {https://ui.adsabs.harvard.edu/abs/2016AJ....151...44D}
}

@ARTICLE{Dawson2013,
       author = {{Dawson}, Kyle S. and {Schlegel}, David J. and {Ahn}, Christopher P. and {Anderson}, Scott F. and {Aubourg}, {\'E}ric and {Bailey}, Stephen and {Barkhouser}, Robert H. and {Bautista}, Julian E. and {Beifiori}, Alessandra and {Berlind}, Andreas A. and {Bhardwaj}, Vaishali and {Bizyaev}, Dmitry and {Blake}, Cullen H. and {Blanton}, Michael R. and {Blomqvist}, Michael and {Bolton}, Adam S. and {Borde}, Arnaud and {Bovy}, Jo and {Brandt}, W.~N. and {Brewington}, Howard and {Brinkmann}, Jon and {Brown}, Peter J. and {Brownstein}, Joel R. and {Bundy}, Kevin and {Busca}, N.~G. and {Carithers}, William and {Carnero}, Aurelio R. and {Carr}, Michael A. and {Chen}, Yanmei and {Comparat}, Johan and {Connolly}, Natalia and {Cope}, Frances and {Croft}, Rupert A.~C. and {Cuesta}, Antonio J. and {da Costa}, Luiz N. and {Davenport}, James R.~A. and {Delubac}, Timoth{\'e}e and {de Putter}, Roland and {Dhital}, Saurav and {Ealet}, Anne and {Ebelke}, Garrett L. and {Eisenstein}, Daniel J. and {Escoffier}, S. and {Fan}, Xiaohui and {Filiz Ak}, N. and {Finley}, Hayley and {Font-Ribera}, Andreu and {G{\'e}nova-Santos}, R. and {Gunn}, James E. and {Guo}, Hong and {Haggard}, Daryl and {Hall}, Patrick B. and {Hamilton}, Jean-Christophe and {Harris}, Ben and {Harris}, David W. and {Ho}, Shirley and {Hogg}, David W. and {Holder}, Diana and {Honscheid}, Klaus and {Huehnerhoff}, Joe and {Jordan}, Beatrice and {Jordan}, Wendell P. and {Kauffmann}, Guinevere and {Kazin}, Eyal A. and {Kirkby}, David and {Klaene}, Mark A. and {Kneib}, Jean-Paul and {Le Goff}, Jean-Marc and {Lee}, Khee-Gan and {Long}, Daniel C. and {Loomis}, Craig P. and {Lundgren}, Britt and {Lupton}, Robert H. and {Maia}, Marcio A.~G. and {Makler}, Martin and {Malanushenko}, Elena and {Malanushenko}, Viktor and {Mandelbaum}, Rachel and {Manera}, Marc and {Maraston}, Claudia and {Margala}, Daniel and {Masters}, Karen L. and {McBride}, Cameron K. and {McDonald}, Patrick and {McGreer}, Ian D. and {McMahon}, Richard G. and {Mena}, Olga and {Miralda-Escud{\'e}}, Jordi and {Montero-Dorta}, Antonio D. and {Montesano}, Francesco and {Muna}, Demitri and {Myers}, Adam D. and {Naugle}, Tracy and {Nichol}, Robert C. and {Noterdaeme}, Pasquier and {Nuza}, Sebasti{\'a}n E. and {Olmstead}, Matthew D. and {Oravetz}, Audrey and {Oravetz}, Daniel J. and {Owen}, Russell and {Padmanabhan}, Nikhil and {Palanque-Delabrouille}, Nathalie and {Pan}, Kaike and {Parejko}, John K. and {P{\^a}ris}, Isabelle and {Percival}, Will J. and {P{\'e}rez-Fournon}, Ismael and {P{\'e}rez-R{\`a}fols}, Ignasi and {Petitjean}, Patrick and {Pfaffenberger}, Robert and {Pforr}, Janine and {Pieri}, Matthew M. and {Prada}, Francisco and {Price-Whelan}, Adrian M. and {Raddick}, M. Jordan and {Rebolo}, Rafael and {Rich}, James and {Richards}, Gordon T. and {Rockosi}, Constance M. and {Roe}, Natalie A. and {Ross}, Ashley J. and {Ross}, Nicholas P. and {Rossi}, Graziano and {Rubi{\~n}o-Martin}, J.~A. and {Samushia}, Lado and {S{\'a}nchez}, Ariel G. and {Sayres}, Conor and {Schmidt}, Sarah J. and {Schneider}, Donald P. and {Sc{\'o}ccola}, C.~G. and {Seo}, Hee-Jong and {Shelden}, Alaina and {Sheldon}, Erin and {Shen}, Yue and {Shu}, Yiping and {Slosar}, An{\v{z}}e and {Smee}, Stephen A. and {Snedden}, Stephanie A. and {Stauffer}, Fritz and {Steele}, Oliver and {Strauss}, Michael A. and {Streblyanska}, Alina and {Suzuki}, Nao and {Swanson}, Molly E.~C. and {Tal}, Tomer and {Tanaka}, Masayuki and {Thomas}, Daniel and {Tinker}, Jeremy L. and {Tojeiro}, Rita and {Tremonti}, Christy A. and {Vargas Maga{\~n}a}, M. and {Verde}, Licia and {Viel}, Matteo and {Wake}, David A. and {Watson}, Mike and {Weaver}, Benjamin A. and {Weinberg}, David H. and {Weiner}, Benjamin J. and {West}, Andrew A. and {White}, Martin and {Wood-Vasey}, W.~M. and {Yeche}, Christophe and {Zehavi}, Idit and {Zhao}, Gong-Bo and {Zheng}, Zheng},
        title = "{The Baryon Oscillation Spectroscopic Survey of SDSS-III}",
      journal = {\aj},
         year = 2013,
        month = jan,
       volume = {145},
       number = {1},
          eid = {10},
        pages = {10},
          doi = {10.1088/0004-6256/145/1/10},
archivePrefix = {arXiv},
       eprint = {1208.0022},
 primaryClass = {astro-ph.CO},
       adsurl = {https://ui.adsabs.harvard.edu/abs/2013AJ....145...10D}
}

@ARTICLE{Eisenstein2005,
       author = {{Eisenstein}, Daniel J. and {Zehavi}, Idit and {Hogg}, David W. and {Scoccimarro}, Roman and {Blanton}, Michael R. and {Nichol}, Robert C. and {Scranton}, Ryan and {Seo}, Hee-Jong and {Tegmark}, Max and {Zheng}, Zheng and {Anderson}, Scott F. and {Annis}, Jim and {Bahcall}, Neta and {Brinkmann}, Jon and {Burles}, Scott and {Castander}, Francisco J. and {Connolly}, Andrew and {Csabai}, Istvan and {Doi}, Mamoru and {Fukugita}, Masataka and {Frieman}, Joshua A. and {Glazebrook}, Karl and {Gunn}, James E. and {Hendry}, John S. and {Hennessy}, Gregory and {Ivezi{\'c}}, Zeljko and {Kent}, Stephen and {Knapp}, Gillian R. and {Lin}, Huan and {Loh}, Yeong-Shang and {Lupton}, Robert H. and {Margon}, Bruce and {McKay}, Timothy A. and {Meiksin}, Avery and {Munn}, Jeffery A. and {Pope}, Adrian and {Richmond}, Michael W. and {Schlegel}, David and {Schneider}, Donald P. and {Shimasaku}, Kazuhiro and {Stoughton}, Christopher and {Strauss}, Michael A. and {SubbaRao}, Mark and {Szalay}, Alexander S. and {Szapudi}, Istv{\'a}n and {Tucker}, Douglas L. and {Yanny}, Brian and {York}, Donald G.},
        title = "{Detection of the Baryon Acoustic Peak in the Large-Scale Correlation Function of SDSS Luminous Red Galaxies}",
      journal = {\apj},
         year = 2005,
        month = nov,
       volume = {633},
       number = {2},
        pages = {560-574},
          doi = {10.1086/466512},
archivePrefix = {arXiv},
       eprint = {astro-ph/0501171},
 primaryClass = {astro-ph},
       adsurl = {https://ui.adsabs.harvard.edu/abs/2005ApJ...633..560E}
}

@ARTICLE{Cole2005,
       author = {{Cole}, Shaun and {Percival}, Will J. and {Peacock}, John A. and {Norberg}, Peder and {Baugh}, Carlton M. and {Frenk}, Carlos S. and {Baldry}, Ivan and {Bland-Hawthorn}, Joss and {Bridges}, Terry and {Cannon}, Russell and {Colless}, Matthew and {Collins}, Chris and {Couch}, Warrick and {Cross}, Nicholas J.~G. and {Dalton}, Gavin and {Eke}, Vincent R. and {De Propris}, Roberto and {Driver}, Simon P. and {Efstathiou}, George and {Ellis}, Richard S. and {Glazebrook}, Karl and {Jackson}, Carole and {Jenkins}, Adrian and {Lahav}, Ofer and {Lewis}, Ian and {Lumsden}, Stuart and {Maddox}, Steve and {Madgwick}, Darren and {Peterson}, Bruce A. and {Sutherland}, Will and {Taylor}, Keith},
        title = "{The 2dF Galaxy Redshift Survey: power-spectrum analysis of the final data set and cosmological implications}",
      journal = {\mnras},
         year = 2005,
        month = sep,
       volume = {362},
       number = {2},
        pages = {505-534},
          doi = {10.1111/j.1365-2966.2005.09318.x},
archivePrefix = {arXiv},
       eprint = {astro-ph/0501174},
 primaryClass = {astro-ph},
       adsurl = {https://ui.adsabs.harvard.edu/abs/2005MNRAS.362..505C}
}

@ARTICLE{Han16,
       author = {{Han}, Jiaxin and {Cole}, Shaun and {Frenk}, Carlos S. and {Jing}, Yipeng},
        title = "{A unified model for the spatial and mass distribution of subhaloes}",
      journal = {\mnras},
         year = 2016,
        month = apr,
       volume = {457},
       number = {2},
        pages = {1208-1223},
          doi = {10.1093/mnras/stv2900},
archivePrefix = {arXiv},
       eprint = {1509.02175},
 primaryClass = {astro-ph.CO},
       adsurl = {https://ui.adsabs.harvard.edu/abs/2016MNRAS.457.1208H}
}

@ARTICLE{Jiang2026,
       author = {{Jiang}, Wenkang and {Han}, Jiaxin and {Xu}, Kun and {Forouhar Moreno}, Victor J. and {He}, Feihong and {Li}, Zhaozhou and {Jiang}, Chunyan and {Jing}, Yipeng and {Yang}, Xiaohu},
        title = "{The Broken Similarity: Sinking and Merging of Dark Matter Subhalos Across Hierarchical Levels}",
      journal = {arXiv e-prints},
         year = 2026,
        month = apr,
          eid = {arXiv:2604.05580},
        pages = {arXiv:2604.05580},
          doi = {10.48550/arXiv.2604.05580},
archivePrefix = {arXiv},
       eprint = {2604.05580},
 primaryClass = {astro-ph.GA},
       adsurl = {https://ui.adsabs.harvard.edu/abs/2026arXiv260405580J}
}

@ARTICLE{Armijo2026,
       author = {{Armijo}, Joaquin and {Da Costa}, Lucas},
        title = "{Marked statistics across the cosmic web: Environmental dependent clustering in modified gravity simulations}",
      journal = {arXiv e-prints},
         year = 2026,
        month = jan,
          eid = {arXiv:2601.05934},
        pages = {arXiv:2601.05934},
          doi = {10.48550/arXiv.2601.05934},
archivePrefix = {arXiv},
       eprint = {2601.05934},
 primaryClass = {astro-ph.CO},
       adsurl = {https://ui.adsabs.harvard.edu/abs/2026arXiv260105934A}
}

@ARTICLE{Wang2025,
       author = {{Wang}, Yirong and {Zhai}, Zhongxu and {Yang}, Xiaohu and {Tinker}, Jeremy L.},
        title = "{Exploring the Signature of Assembly Bias and Modified Gravity Using Small-scale Clusterings of Galaxies}",
      journal = {\apj},
         year = 2025,
        month = nov,
       volume = {994},
       number = {1},
          eid = {51},
        pages = {51},
          doi = {10.3847/1538-4357/ae0c10},
archivePrefix = {arXiv},
       eprint = {2506.22737},
 primaryClass = {astro-ph.CO},
       adsurl = {https://ui.adsabs.harvard.edu/abs/2025ApJ...994...51W}
}

@ARTICLE{Jiang2025,
       author = {{Jiang}, Wenkang and {Han}, Jiaxin and {Dong}, Fuyu and {He}, Feihong},
        title = "{Self-similar Decomposition of the Hierarchical Merger Tree of Dark Matter Halos}",
      journal = {\apj},
         year = 2025,
        month = aug,
       volume = {988},
       number = {2},
          eid = {160},
        pages = {160},
          doi = {10.3847/1538-4357/ade439},
archivePrefix = {arXiv},
       eprint = {2502.20181},
 primaryClass = {astro-ph.CO},
       adsurl = {https://ui.adsabs.harvard.edu/abs/2025ApJ...988..160J}
}

@ARTICLE{Dong2022,
       author = {{Dong}, Fuyu and {Zhao}, Donghai and {Han}, Jiaxin and {Li}, Zhaozhou and {Jing}, Yipeng and {Yang}, Xiaohu},
        title = "{The Universal Specific Merger Rate of Dark Matter Halos}",
      journal = {\apj},
         year = 2022,
        month = apr,
       volume = {929},
       number = {2},
          eid = {120},
        pages = {120},
          doi = {10.3847/1538-4357/ac5aaa},
archivePrefix = {arXiv},
       eprint = {2112.08047},
 primaryClass = {astro-ph.CO},
       adsurl = {https://ui.adsabs.harvard.edu/abs/2022ApJ...929..120D}
}

@ARTICLE{Gupta2022,
       author = {{Gupta}, Suhani and {Hellwing}, Wojciech A. and {Bilicki}, Maciej and {Garc{\'\i}a-Farieta}, Jorge Enrique},
        title = "{Universality of the halo mass function in modified gravity cosmologies}",
      journal = {\prd},
         year = 2022,
        month = feb,
       volume = {105},
       number = {4},
          eid = {043538},
        pages = {043538},
          doi = {10.1103/PhysRevD.105.043538},
archivePrefix = {arXiv},
       eprint = {2112.03699},
 primaryClass = {astro-ph.CO},
       adsurl = {https://ui.adsabs.harvard.edu/abs/2022PhRvD.105d3538G}
}

@ARTICLE{Ishak2019,
       author = {{Ishak}, Mustapha},
        title = "{Testing general relativity in cosmology}",
      journal = {Living Reviews in Relativity},
         year = 2019,
        month = dec,
       volume = {22},
       number = {1},
          eid = {1},
        pages = {1},
          doi = {10.1007/s41114-018-0017-4},
archivePrefix = {arXiv},
       eprint = {1806.10122},
 primaryClass = {astro-ph.CO},
       adsurl = {https://ui.adsabs.harvard.edu/abs/2019LRR....22....1I}
}

@ARTICLE{Hernandez-Aguayo2019,
       author = {{Hern{\'a}ndez-Aguayo}, C{\'e}sar and {Hou}, Jiamin and {Li}, Baojiu and {Baugh}, Carlton M. and {S{\'a}nchez}, Ariel G.},
        title = "{Large-scale redshift space distortions in modified gravity theories}",
      journal = {\mnras},
         year = 2019,
        month = may,
       volume = {485},
       number = {2},
        pages = {2194-2213},
          doi = {10.1093/mnras/stz516},
archivePrefix = {arXiv},
       eprint = {1811.09197},
 primaryClass = {astro-ph.CO},
       adsurl = {https://ui.adsabs.harvard.edu/abs/2019MNRAS.485.2194H}
}

@ARTICLE{Paillas2019,
       author = {{Paillas}, Enrique and {Cautun}, Marius and {Li}, Baojiu and {Cai}, Yan-Chuan and {Padilla}, Nelson and {Armijo}, Joaqu{\'\i}n and {Bose}, Sownak},
        title = "{The Santiago-Harvard-Edinburgh-Durham void comparison II: unveiling the Vainshtein screening using weak lensing}",
      journal = {\mnras},
         year = 2019,
        month = mar,
       volume = {484},
       number = {1},
        pages = {1149-1165},
          doi = {10.1093/mnras/stz022},
archivePrefix = {arXiv},
       eprint = {1810.02864},
 primaryClass = {astro-ph.CO},
       adsurl = {https://ui.adsabs.harvard.edu/abs/2019MNRAS.484.1149P}
}

@ARTICLE{Hernandez-Aguayo2018,
       author = {{Hern{\'a}ndez-Aguayo}, C{\'e}sar and {Baugh}, Carlton M. and {Li}, Baojiu},
        title = "{Marked clustering statistics in f(R) gravity cosmologies}",
      journal = {\mnras},
         year = 2018,
        month = oct,
       volume = {479},
       number = {4},
        pages = {4824-4835},
          doi = {10.1093/mnras/sty1822},
archivePrefix = {arXiv},
       eprint = {1801.08880},
 primaryClass = {astro-ph.CO},
       adsurl = {https://ui.adsabs.harvard.edu/abs/2018MNRAS.479.4824H}
}

@ARTICLE{Armijo2018,
       author = {{Armijo}, Joaqu{\'\i}n and {Cai}, Yan-Chuan and {Padilla}, Nelson and {Li}, Baojiu and {Peacock}, John A.},
        title = "{Testing modified gravity using a marked correlation function}",
      journal = {\mnras},
         year = 2018,
        month = aug,
       volume = {478},
       number = {3},
        pages = {3627-3632},
          doi = {10.1093/mnras/sty1335},
archivePrefix = {arXiv},
       eprint = {1801.08975},
 primaryClass = {astro-ph.CO},
       adsurl = {https://ui.adsabs.harvard.edu/abs/2018MNRAS.478.3627A}
}

@ARTICLE{Cautun2018,
       author = {{Cautun}, Marius and {Paillas}, Enrique and {Cai}, Yan-Chuan and {Bose}, Sownak and {Armijo}, Joaquin and {Li}, Baojiu and {Padilla}, Nelson},
        title = "{The Santiago-Harvard-Edinburgh-Durham void comparison - I. SHEDding light on chameleon gravity tests}",
      journal = {\mnras},
         year = 2018,
        month = may,
       volume = {476},
       number = {3},
        pages = {3195-3217},
          doi = {10.1093/mnras/sty463},
archivePrefix = {arXiv},
       eprint = {1710.01730},
 primaryClass = {astro-ph.CO},
       adsurl = {https://ui.adsabs.harvard.edu/abs/2018MNRAS.476.3195C}
}

@ARTICLE{Han2018,
       author = {{Han}, Jiaxin and {Cole}, Shaun and {Frenk}, Carlos S. and {Benitez-Llambay}, Alejandro and {Helly}, John},
        title = "{HBT+: an improved code for finding subhaloes and building merger trees in cosmological simulations}",
      journal = {\mnras},
         year = 2018,
        month = feb,
       volume = {474},
       number = {1},
        pages = {604-617},
          doi = {10.1093/mnras/stx2792},
archivePrefix = {arXiv},
       eprint = {1708.03646},
 primaryClass = {astro-ph.CO},
       adsurl = {https://ui.adsabs.harvard.edu/abs/2018MNRAS.474..604H}
}

@ARTICLE{Poole2017,
       author = {{Poole}, Gregory B. and {Mutch}, Simon J. and {Croton}, Darren J. and {Wyithe}, Stuart},
        title = "{Convergence properties of halo merger trees; halo and substructure merger rates across cosmic history}",
      journal = {\mnras},
         year = 2017,
        month = dec,
       volume = {472},
       number = {3},
        pages = {3659-3682},
          doi = {10.1093/mnras/stx2233},
archivePrefix = {arXiv},
       eprint = {1709.01526},
 primaryClass = {astro-ph.GA},
       adsurl = {https://ui.adsabs.harvard.edu/abs/2017MNRAS.472.3659P}
}

@ARTICLE{Arnalte-Mur2017,
       author = {{Arnalte-Mur}, Pablo and {Hellwing}, Wojciech A. and {Norberg}, Peder},
        title = "{Real- and redshift-space halo clustering in f(R) cosmologies}",
      journal = {\mnras},
         year = 2017,
        month = may,
       volume = {467},
       number = {2},
        pages = {1569-1585},
          doi = {10.1093/mnras/stx196},
archivePrefix = {arXiv},
       eprint = {1612.02355},
 primaryClass = {astro-ph.CO},
       adsurl = {https://ui.adsabs.harvard.edu/abs/2017MNRAS.467.1569A}
}

@ARTICLE{Koyama2016,
       author = {{Koyama}, Kazuya},
        title = "{Cosmological tests of modified gravity}",
      journal = {Reports on Progress in Physics},
         year = 2016,
        month = apr,
       volume = {79},
       number = {4},
          eid = {046902},
        pages = {046902},
          doi = {10.1088/0034-4885/79/4/046902},
archivePrefix = {arXiv},
       eprint = {1504.04623},
 primaryClass = {astro-ph.CO},
       adsurl = {https://ui.adsabs.harvard.edu/abs/2016RPPh...79d6902K}
}

@ARTICLE{Gronke2015,
       author = {{Gronke}, M. and {Llinares}, C. and {Mota}, D.~F. and {Winther}, H.~A.},
        title = "{Halo velocity profiles in screened modified gravity theories}",
      journal = {\mnras},
         year = 2015,
        month = may,
       volume = {449},
       number = {3},
        pages = {2837-2844},
          doi = {10.1093/mnras/stv496},
archivePrefix = {arXiv},
       eprint = {1412.0066},
 primaryClass = {astro-ph.CO},
       adsurl = {https://ui.adsabs.harvard.edu/abs/2015MNRAS.449.2837G}
}

@ARTICLE{Moran2015,
       author = {{Moran}, C. Corbett and {Teyssier}, R. and {Li}, B.},
        title = "{Chameleon f(R) gravity on the Virgo cluster scale}",
      journal = {\mnras},
         year = 2015,
        month = mar,
       volume = {448},
       number = {1},
        pages = {307-327},
          doi = {10.1093/mnras/stu2757},
       adsurl = {https://ui.adsabs.harvard.edu/abs/2015MNRAS.448..307C}
}

@ARTICLE{Li2013,
       author = {{Li}, Baojiu and {Zhao}, Gong-Bo and {Koyama}, Kazuya},
        title = "{Exploring Vainshtein mechanism on adaptively refined meshes}",
      journal = {\jcap},
         year = 2013,
        month = may,
       volume = {2013},
       number = {5},
          eid = {023},
        pages = {023},
          doi = {10.1088/1475-7516/2013/05/023},
archivePrefix = {arXiv},
       eprint = {1303.0008},
 primaryClass = {astro-ph.CO},
       adsurl = {https://ui.adsabs.harvard.edu/abs/2013JCAP...05..023L}
}

@ARTICLE{Han2012,
       author = {{Han}, Jiaxin and {Jing}, Y.~P. and {Wang}, Huiyuan and {Wang}, Wenting},
        title = "{Resolving subhaloes' lives with the Hierarchical Bound-Tracing algorithm}",
      journal = {\mnras},
         year = 2012,
        month = dec,
       volume = {427},
       number = {3},
        pages = {2437-2449},
          doi = {10.1111/j.1365-2966.2012.22111.x},
archivePrefix = {arXiv},
       eprint = {1103.2099},
 primaryClass = {astro-ph.CO},
       adsurl = {https://ui.adsabs.harvard.edu/abs/2012MNRAS.427.2437H}
}

@ARTICLE{Li2012a,
       author = {{Li}, Baojiu and {Zhao}, Gong-Bo and {Koyama}, Kazuya},
        title = "{Haloes and voids in f(R) gravity}",
      journal = {\mnras},
         year = 2012,
        month = apr,
       volume = {421},
       number = {4},
        pages = {3481-3487},
          doi = {10.1111/j.1365-2966.2012.20573.x},
archivePrefix = {arXiv},
       eprint = {1111.2602},
 primaryClass = {astro-ph.CO},
       adsurl = {https://ui.adsabs.harvard.edu/abs/2012MNRAS.421.3481L}
}

@ARTICLE{Clifton2012,
       author = {{Clifton}, Timothy and {Ferreira}, Pedro G. and {Padilla}, Antonio and {Skordis}, Constantinos},
        title = "{Modified gravity and cosmology}",
      journal = {\physrep},
         year = 2012,
        month = mar,
       volume = {513},
       number = {1},
        pages = {1-189},
          doi = {10.1016/j.physrep.2012.01.001},
archivePrefix = {arXiv},
       eprint = {1106.2476},
 primaryClass = {astro-ph.CO},
       adsurl = {https://ui.adsabs.harvard.edu/abs/2012PhR...513....1C}
}

@ARTICLE{Li2012b,
       author = {{Li}, Baojiu and {Zhao}, Gong-Bo and {Teyssier}, Romain and {Koyama}, Kazuya},
        title = "{ECOSMOG: an Efficient COde for Simulating MOdified Gravity}",
      journal = {\jcap},
         year = 2012,
        month = jan,
       volume = {2012},
       number = {1},
          eid = {051},
        pages = {051},
          doi = {10.1088/1475-7516/2012/01/051},
archivePrefix = {arXiv},
       eprint = {1110.1379},
 primaryClass = {astro-ph.CO},
       adsurl = {https://ui.adsabs.harvard.edu/abs/2012JCAP...01..051L}
}

@ARTICLE{DeFelice2010,
       author = {{De Felice}, Antonio and {Tsujikawa}, Shinji},
        title = "{f( R) Theories}",
      journal = {Living Reviews in Relativity},
         year = 2010,
        month = dec,
       volume = {13},
       number = {1},
          eid = {3},
        pages = {3},
          doi = {10.12942/lrr-2010-3},
archivePrefix = {arXiv},
       eprint = {1002.4928},
 primaryClass = {gr-qc},
       adsurl = {https://ui.adsabs.harvard.edu/abs/2010LRR....13....3D}
}

@ARTICLE{Maartens2010,
       author = {{Maartens}, Roy and {Koyama}, Kazuya},
        title = "{Brane-World Gravity}",
      journal = {Living Reviews in Relativity},
         year = 2010,
        month = dec,
       volume = {13},
       number = {1},
          eid = {5},
        pages = {5},
          doi = {10.12942/lrr-2010-5},
archivePrefix = {arXiv},
       eprint = {1004.3962},
 primaryClass = {hep-th},
       adsurl = {https://ui.adsabs.harvard.edu/abs/2010LRR....13....5M}
}

@ARTICLE{Genel2010,
       author = {{Genel}, Shy and {Bouch{\'e}}, Nicolas and {Naab}, Thorsten and {Sternberg}, Amiel and {Genzel}, Reinhard},
        title = "{The Growth of Dark Matter Halos: Evidence for Significant Smooth Accretion}",
      journal = {\apj},
         year = 2010,
        month = aug,
       volume = {719},
       number = {1},
        pages = {229-239},
          doi = {10.1088/0004-637X/719/1/229},
archivePrefix = {arXiv},
       eprint = {1005.4058},
 primaryClass = {astro-ph.CO},
       adsurl = {https://ui.adsabs.harvard.edu/abs/2010ApJ...719..229G}
}

@ARTICLE{Fakhouri2010,
       author = {{Fakhouri}, Onsi and {Ma}, Chung-Pei and {Boylan-Kolchin}, Michael},
        title = "{The merger rates and mass assembly histories of dark matter haloes in the two Millennium simulations}",
      journal = {\mnras},
         year = 2010,
        month = aug,
       volume = {406},
       number = {4},
        pages = {2267-2278},
          doi = {10.1111/j.1365-2966.2010.16859.x},
archivePrefix = {arXiv},
       eprint = {1001.2304},
 primaryClass = {astro-ph.CO},
       adsurl = {https://ui.adsabs.harvard.edu/abs/2010MNRAS.406.2267F}
}

@ARTICLE{Sotiriou2010,
       author = {{Sotiriou}, Thomas P. and {Faraoni}, Valerio},
        title = "{f(R) theories of gravity}",
      journal = {Reviews of Modern Physics},
         year = 2010,
        month = jan,
       volume = {82},
       number = {1},
        pages = {451-497},
          doi = {10.1103/RevModPhys.82.451},
archivePrefix = {arXiv},
       eprint = {0805.1726},
 primaryClass = {gr-qc},
       adsurl = {https://ui.adsabs.harvard.edu/abs/2010RvMP...82..451S}
}

@ARTICLE{McBride2009,
       author = {{McBride}, James and {Fakhouri}, Onsi and {Ma}, Chung-Pei},
        title = "{Mass accretion rates and histories of dark matter haloes}",
      journal = {\mnras},
         year = 2009,
        month = oct,
       volume = {398},
       number = {4},
        pages = {1858-1868},
          doi = {10.1111/j.1365-2966.2009.15329.x},
archivePrefix = {arXiv},
       eprint = {0902.3659},
 primaryClass = {astro-ph.CO},
       adsurl = {https://ui.adsabs.harvard.edu/abs/2009MNRAS.398.1858M}
}

@ARTICLE{Stewart2009,
       author = {{Stewart}, Kyle R. and {Bullock}, James S. and {Barton}, Elizabeth J. and {Wechsler}, Risa H.},
        title = "{Galaxy Mergers and Dark Matter Halo Mergers in {\ensuremath{\Lambda}}CDM: Mass, Redshift, and Mass-Ratio Dependence}",
      journal = {\apj},
         year = 2009,
        month = sep,
       volume = {702},
       number = {2},
        pages = {1005-1015},
          doi = {10.1088/0004-637X/702/2/1005},
archivePrefix = {arXiv},
       eprint = {0811.1218},
 primaryClass = {astro-ph},
       adsurl = {https://ui.adsabs.harvard.edu/abs/2009ApJ...702.1005S}
}

@ARTICLE{Genel2009,
       author = {{Genel}, Shy and {Genzel}, Reinhard and {Bouch{\'e}}, Nicolas and {Naab}, Thorsten and {Sternberg}, Amiel},
        title = "{The Halo Merger Rate in the Millennium Simulation and Implications for Observed Galaxy Merger Fractions}",
      journal = {\apj},
         year = 2009,
        month = aug,
       volume = {701},
       number = {2},
        pages = {2002-2018},
          doi = {10.1088/0004-637X/701/2/2002},
archivePrefix = {arXiv},
       eprint = {0812.3154},
 primaryClass = {astro-ph},
       adsurl = {https://ui.adsabs.harvard.edu/abs/2009ApJ...701.2002G}
}

@ARTICLE{Wetzel2009,
       author = {{Wetzel}, Andrew R. and {Cohn}, J.~D. and {White}, Martin},
        title = "{Simulating subhaloes at high redshift: merger rates, counts and types}",
      journal = {\mnras},
         year = 2009,
        month = may,
       volume = {395},
       number = {3},
        pages = {1376-1390},
          doi = {10.1111/j.1365-2966.2009.14424.x},
archivePrefix = {arXiv},
       eprint = {0810.2537},
 primaryClass = {astro-ph},
       adsurl = {https://ui.adsabs.harvard.edu/abs/2009MNRAS.395.1376W}
}

@ARTICLE{Fakhouri2009,
       author = {{Fakhouri}, Onsi and {Ma}, Chung-Pei},
        title = "{Environmental dependence of dark matter halo growth - I. Halo merger rates}",
      journal = {\mnras},
         year = 2009,
        month = apr,
       volume = {394},
       number = {4},
        pages = {1825-1840},
          doi = {10.1111/j.1365-2966.2009.14480.x},
archivePrefix = {arXiv},
       eprint = {0808.2471},
 primaryClass = {astro-ph},
       adsurl = {https://ui.adsabs.harvard.edu/abs/2009MNRAS.394.1825F}
}

@ARTICLE{Prunet2008,
       author = {{Prunet}, S. and {Pichon}, C. and {Aubert}, D. and {Pogosyan}, D. and {Teyssier}, R. and {Gottloeber}, S.},
        title = "{Initial Conditions For Large Cosmological Simulations}",
      journal = {\apjs},
         year = 2008,
        month = oct,
       volume = {178},
       number = {2},
        pages = {179-188},
          doi = {10.1086/590370},
archivePrefix = {arXiv},
       eprint = {0804.3536},
 primaryClass = {astro-ph},
       adsurl = {https://ui.adsabs.harvard.edu/abs/2008ApJS..178..179P}
}

@ARTICLE{Fakhouri2008,
       author = {{Fakhouri}, Onsi and {Ma}, Chung-Pei},
        title = "{The nearly universal merger rate of dark matter haloes in {\ensuremath{\Lambda}}CDM cosmology}",
      journal = {\mnras},
         year = 2008,
        month = may,
       volume = {386},
       number = {2},
        pages = {577-592},
          doi = {10.1111/j.1365-2966.2008.13075.x},
archivePrefix = {arXiv},
       eprint = {0710.4567},
 primaryClass = {astro-ph},
       adsurl = {https://ui.adsabs.harvard.edu/abs/2008MNRAS.386..577F}
}

@ARTICLE{Hu2007,
       author = {{Hu}, Wayne and {Sawicki}, Ignacy},
        title = "{Models of f(R) cosmic acceleration that evade solar system tests}",
      journal = {\prd},
         year = 2007,
        month = sep,
       volume = {76},
       number = {6},
          eid = {064004},
        pages = {064004},
          doi = {10.1103/PhysRevD.76.064004},
archivePrefix = {arXiv},
       eprint = {0705.1158},
 primaryClass = {astro-ph},
       adsurl = {https://ui.adsabs.harvard.edu/abs/2007PhRvD..76f4004H}
}

@ARTICLE{Copeland2006,
       author = {{Copeland}, Edmund J. and {Sami}, M. and {Tsujikawa}, Shinji},
        title = "{Dynamics of Dark Energy}",
      journal = {International Journal of Modern Physics D},
         year = 2006,
        month = jan,
       volume = {15},
       number = {11},
        pages = {1753-1935},
          doi = {10.1142/S021827180600942X},
archivePrefix = {arXiv},
       eprint = {hep-th/0603057},
 primaryClass = {hep-th},
       adsurl = {https://ui.adsabs.harvard.edu/abs/2006IJMPD..15.1753C}
}

@ARTICLE{Khoury2004,
       author = {{Khoury}, Justin and {Weltman}, Amanda},
        title = "{Chameleon cosmology}",
      journal = {\prd},
         year = 2004,
        month = feb,
       volume = {69},
       number = {4},
          eid = {044026},
        pages = {044026},
          doi = {10.1103/PhysRevD.69.044026},
archivePrefix = {arXiv},
       eprint = {astro-ph/0309411},
 primaryClass = {astro-ph},
       adsurl = {https://ui.adsabs.harvard.edu/abs/2004PhRvD..69d4026K}
}

@ARTICLE{Dvali2000,
       author = {{Dvali}, G. and {Gabadadze}, G. and {Porrati}, M.},
        title = "{4D gravity on a brane in 5D Minkowski space}",
      journal = {Physics Letters B},
         year = 2000,
        month = jul,
       volume = {485},
       number = {1-3},
        pages = {208-214},
          doi = {10.1016/S0370-2693(00)00669-9},
archivePrefix = {arXiv},
       eprint = {hep-th/0005016},
 primaryClass = {hep-th},
       adsurl = {https://ui.adsabs.harvard.edu/abs/2000PhLB..485..208D}
}

@ARTICLE{Perlmutter1999,
       author = {{Perlmutter}, S. and {Aldering}, G. and {Goldhaber}, G. and {Knop}, R.~A. and {Nugent}, P. and {Castro}, P.~G. and {Deustua}, S. and {Fabbro}, S. and {Goobar}, A. and {Groom}, D.~E. and {Hook}, I.~M. and {Kim}, A.~G. and {Kim}, M.~Y. and {Lee}, J.~C. and {Nunes}, N.~J. and {Pain}, R. and {Pennypacker}, C.~R. and {Quimby}, R. and {Lidman}, C. and {Ellis}, R.~S. and {Irwin}, M. and {McMahon}, R.~G. and {Ruiz-Lapuente}, P. and {Walton}, N. and {Schaefer}, B. and {Boyle}, B.~J. and {Filippenko}, A.~V. and {Matheson}, T. and {Fruchter}, A.~S. and {Panagia}, N. and {Newberg}, H.~J.~M. and {Couch}, W.~J. and {Project}, The Supernova Cosmology},
        title = "{Measurements of {\ensuremath{\Omega}} and {\ensuremath{\Lambda}} from 42 High-Redshift Supernovae}",
      journal = {\apj},
         year = 1999,
        month = jun,
       volume = {517},
       number = {2},
        pages = {565-586},
          doi = {10.1086/307221},
archivePrefix = {arXiv},
       eprint = {astro-ph/9812133},
 primaryClass = {astro-ph},
       adsurl = {https://ui.adsabs.harvard.edu/abs/1999ApJ...517..565P}
}

@ARTICLE{Riess1998,
       author = {{Riess}, Adam G. and {Filippenko}, Alexei V. and {Challis}, Peter and {Clocchiatti}, Alejandro and {Diercks}, Alan and {Garnavich}, Peter M. and {Gilliland}, Ron L. and {Hogan}, Craig J. and {Jha}, Saurabh and {Kirshner}, Robert P. and {Leibundgut}, B. and {Phillips}, M.~M. and {Reiss}, David and {Schmidt}, Brian P. and {Schommer}, Robert A. and {Smith}, R. Chris and {Spyromilio}, J. and {Stubbs}, Christopher and {Suntzeff}, Nicholas B. and {Tonry}, John},
        title = "{Observational Evidence from Supernovae for an Accelerating Universe and a Cosmological Constant}",
      journal = {\aj},
         year = 1998,
        month = sep,
       volume = {116},
       number = {3},
        pages = {1009-1038},
          doi = {10.1086/300499},
archivePrefix = {arXiv},
       eprint = {astro-ph/9805201},
 primaryClass = {astro-ph},
       adsurl = {https://ui.adsabs.harvard.edu/abs/1998AJ....116.1009R}
}

@ARTICLE{Vainshtein1972,
       author = {{Vainshtein}, A.~I.},
        title = "{To the problem of nonvanishing gravitation mass}",
      journal = {Physics Letters B},
         year = 1972,
        month = may,
       volume = {39},
       number = {3},
        pages = {393-394},
          doi = {10.1016/0370-2693(72)90147-5},
       adsurl = {https://ui.adsabs.harvard.edu/abs/1972PhLB...39..393V}
}
\bibliographystyle{aasjournalv7}

\end{document}